# Design of a cw, low energy, high power superconducting linac for environmental applications

G. Ciovati[1], J. Anderson[2], B. Coriton[2], J. Guo[1], F. Hannon[1], L. Holland[2], M. LeSher[2], F. Marhauser[1], J. Rathke[3], R. Rimmer[1], T. Schultheiss[3], V. Vylet[1]

[1]Thomas Jefferson National Accelerator Facility (Jefferson Lab), Newport News, Virginia 23606, USA
[2]General Atomics, San Diego, CA 92186, USA
[3]Advanced Energy Systems, Inc., Medford, New York 11763, USA

**Abstract**

The treatment of flue gases from power plants and municipal or industrial wastewater using electron beam irradiation technology has been successfully demonstrated in small-scale pilot plants. The beam energy requirement is rather modest, on the order of a few MeV, however the adoption of the technology at an industrial scale requires the availability of high beam power, of the order of 1 MW, in a cost effective way. In this article we present the design of a compact superconducting accelerator capable of delivering a cw electron beam with a current of 1 A and an energy of 1 MeV. The main components are an rf-gridded thermionic gun and a conduction cooled $\beta = 0.5$ elliptical $Nb_3Sn$ cavity with dual coaxial power couplers. An engineering and cost analysis shows that the proposed design would result in a processing cost competitive with alternative treatment methods.

List of acronyms

SRF – radio-frequency superconductivity

GM – Gifford-McMahon

BLA – beamline absorber

HOM – higher-order mode

FPC – fundamental power coupler

BBU – beam breakup

YBCO – yttrium barium copper oxide

CW – continuous wave

VED – vacuum electron device

PMA – power module array

MBIOT – multi-beam inductive output tube

NCRP - National Council on Radiation Protection and Measurements

TVL – tenth-value layer

MC – Monte Carlo

EMI – electromagnetic interference

CAPEX – capital expenditure

## I. INTRODUCTION

Electron irradiation is a demonstrated method to reduce contaminants in solid and liquid substances such as flue gases from power plant and wastewater from industries or municipalities [1, 2, 3]. In both cases, the irradiation of gases or liquids with electrons results in the formation of several ions and radicals, which are highly reactive and allow neutralization of contaminants through chemical reactions. In the case of flue gases, electron irradiation along with the injection of ammonia allow eliminating up to ~70% of $NO_x$, ~90 of $SO_x$ and ~99% of Hg as well as removal of volatile organic compounds, with radiation dose in the range 7-12 kGy. $NO_x$ and $SO_x$ are converted to ammonium nitrate and ammonium sulfate, which can be recovered and used as agricultural fertilizer. Irradiation of wastewater at relatively low doses, up to ~1 kGy, has been successfully applied to remove color, odor, decompose organic pollutants and for disinfection. Disinfection of sewage sludge requires higher radiation doses, 4-10 kGy.

In the case of wastewater treatment, the radiation dose is directly proportional to the ratio of the beam power and the mass flow rate. As an example, the pilot plant at a wastewater treatment facility in Miami, Florida showed that a beam power of 75 kW could deliver a dose of ~6.5 kGy to a liquid flow of 460 liter per minute. As another example, the industrial flue gas treatment facility at the Pomorzan electric power station in Szczecin, Poland, used four accelerators with a total power of 1 MW to deliver a dose of 7-12 kGy to flue gases from a 130 MW(e) coal power plant flowing at a maximum rate of 270,000 Normal m$^3$/h.

Typical electron accelerators used in this type of application are dc-type with energies of ~1 MeV and beam powers of few hundreds of kW [3]. To date, the most powerful one is the ELV-12, a cascade accelerator with parallel inductive coupling developed by the Budker Institute of Nuclear Physics in Russia and commercialized by EB Tech Co., Ltd. in South Korea. Such an accelerator has three accelerating tubes providing electrons with energy between 0.6-1 MeV and a total beam power of up to 400 kW, and it has been used in a dyeing wastewater treatment plant in South Korea.

A report published in 2015 by the U.S. Department of Energy assessed existing accelerator-based technologies for energy and environmental applications and provided recommendations for R&D activities aiming at improving the competitiveness of accelerator technology in these sectors [4]. It was discussed in the report that a compact electron accelerator with energy of ~1 MeV and a beam power of at least 1 MW would be a significant advance with respect to accelerators currently available on the market for wastewater and flue gas treatment. In order to be competitive with alternative treatments methods, the treatment cost should be less than 1 $/m$^3$ in the case of wastewater and less than ~100 $ per kW of electricity in the case of flue gases from power plants.

Radio-frequency superconductivity (SRF) is a well-established technology used in high-energy particle accelerators worldwide for research purposes. The main attractiveness of the technology is the ability to provide high accelerating voltages with high efficiency, as the losses in the superconductor are several orders of magnitude smaller than those in normal conductors (copper). The main drawback is the requirement to operate SRF accelerating cavities at liquid He temperature at 2 – 4 K, which involves the installation of large He cryoplants. In this article, we present a design of a compact, cw SRF linear accelerator aiming at delivering an electron beam of energy ~1 MeV and beam power ~1 MW for the treatment of wastewater and flue gases. The layout of the accelerator is presented in Sec. II along with results from beam transport simulations. The design and thermal analysis of the SRF cryomodule are presented in Sec. III. An engineering and cost analysis is discussed in Sec. IV. In Sec. V the results from the design study are discussed along with upgradeability to higher beam energy and further R&D required to lowering the cost.

## II. ACCELERATOR LAYOUT

A schematic layout of the proposed accelerator is shown in Fig. 1. An electron current of 1 A is produced by a gridded thermionic cathode (Fig. 2a) with a cathode-anode potential of 92 kV. A 750 MHz rf signal and a dc bias are applied to the grid to bunch the beam. The low energy beam is focused by a solenoid and accelerated to 1 MeV by a single-cell SRF cavity. An additional solenoid downstream of the cryomodule focuses the beam into the extraction section having a beam scanner and a thin-foil extraction window. Two pumping stations with ion pumps, one between the injector and the cryomodule and one between the cryomodule and the beam scanner, are used to maintain ultra-high vacuum in the beamline.

A gridded thermionic cathode can provide a robust, economical and compact electron source capable of providing high beam current [5]. Cathodes made of IrCe alloy have demonstrated a remarkable lifetime of ~40,000 hours at emission current densities as high as 13 A/cm$^2$ [6]. The industrial application envisioned for this accelerator does not have the stringent electron beam properties often required by the accelerator physics community, so achieving the beam current becomes less constrained. AES has developed a thermionic gun tested up to ~0.8 A at 1 GHz and 23 kV [7]. The current was limited by very short pulse durations, which is not a requirement of this injector. A prototype gridded thermionic gun at the Naval Research Laboratory delivered a peak current of ~2.2 A at 714 MHz and -24 kV cathode voltage [8, 9]. A 300 kV thermionic gun rf-gated at 650 MHz was also recently tested at TRIUMF up to an average current of 10 mA [10].

The cryomodule hosts a 750 MHz SRF single-cell "elliptical" cavity conduction cooled to ~5 K by four Gifford-McMahon (GM) cryocoolers (Fig. 2b). A water-cooled beamline absorber (BLA) such as the one developed at Cornell for CESR [11] is located on one side outside the cryomodule to absorb the power from higher-order modes (HOMs). RF power is coupled into the cavity by two coaxial fundamental power couplers (FPCs) symmetrically located on one side of the cavity.

We assume the extraction device to be very similar to that already developed for the ELV-type accelerators [12]: it features two thin-foil (~50 μm thick) Ti windows, and two electromagnets are used to scan the beam in two mutually orthogonal directions. A switching magnet is used to alternate the beam between the two windows. Both water and air cooling are used to extract heat from the windows region and two vacuum ion pumps are connected directly to the horn.

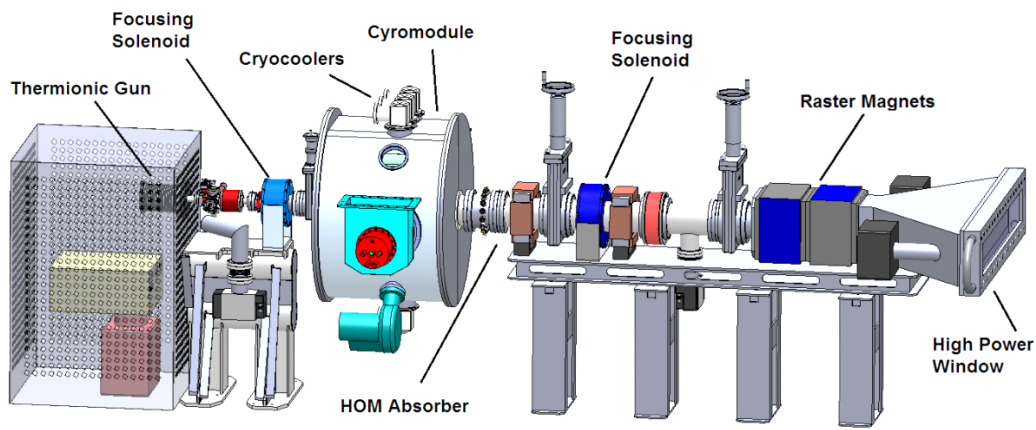

FIG. 1. Schematic layout of the 1 MeV, 1 MW, CW SRF electron accelerator for wastewater and flue gas treatment. The estimated overall length of the accelerator is 6 m.

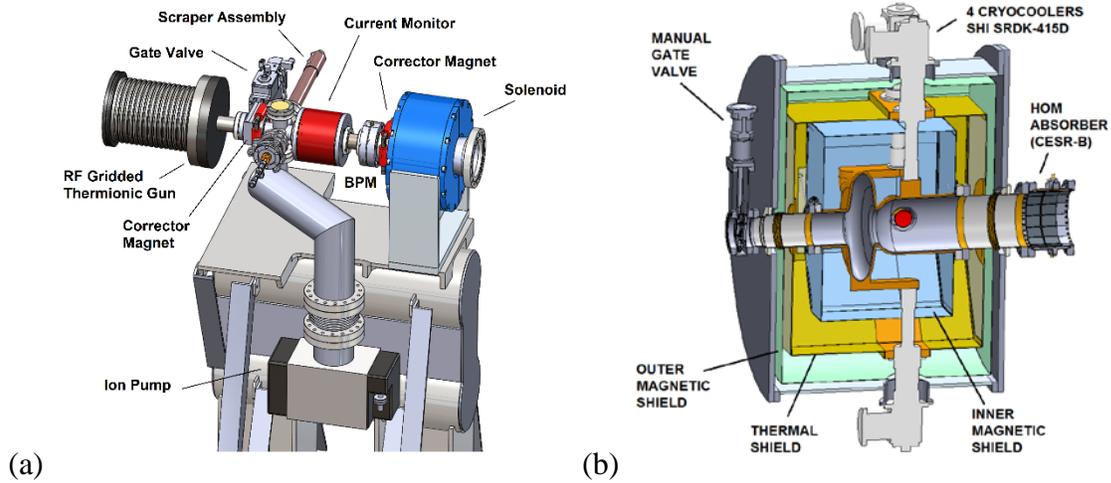

FIG. 2. Close-up views of the injector (a) and the cryomodule assembly (b).

**A. Beam transport simulations**

The proposed beamline has been simulated with the particle tracking software General Particle Tracer [13] using a simplified geometry. The design philosophy was to emulate the operation of the 650 MHz, 100 kV TRIUMF demonstration thermionic gun [9], with the exception of higher current and frequency. Specifically, a commercial gridded dispenser cathode is assumed, but in order to achieve an average current of 1 A (and associated higher peak current), it will have a larger emitting area of 3 cm$^2$ rather than 2 cm$^2$.

The simulated emission from the cathode was assumed to be a truncated Gaussian distribution longitudinally, and radially uniform. To achieve the required current density, the cathode will be required to operate at approximately 2200 K, which results in a calculated thermal emittance of ~6 μm. The thermal emittance was included in the simulation. In the absence of a specific electrode shape inside the thermionic gun, the cathode is assumed to be a flat plate, while the anode is 10 cm downstream modeled by a flat plate with a hole. This gives the on-axis field map shown in Fig. 3.

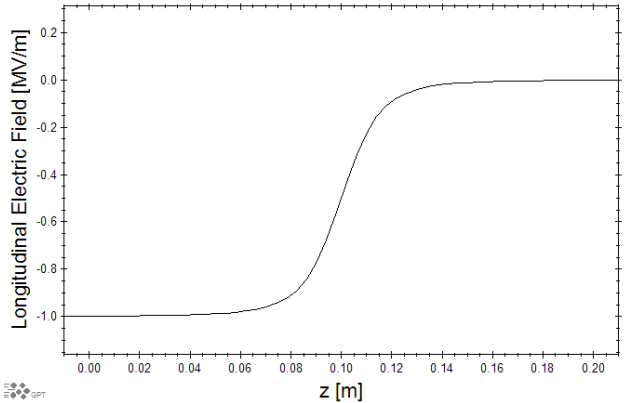

FIG. 3. On-axis longitudinal field profile as a function of distance from the cathode of the simulated electron gun.

Since the beam from the cathode is non-relativistic, a low-$\beta$ ($\beta = v/c$, where $v$ is the particles velocity and $c$ is the speed of light) SRF cavity is required to accelerate the beam. The cavity was initially modeled with a Gaussian on-axis electric field and placed at a location 80 cm downstream of the cathode. A focusing solenoid was placed between the two, 25 cm downstream from the cathode. Downstream of the cavity, at 1.3 m from the cathode, an additional solenoid was used to contain the beam within the beam pipe.

The optimum geometric $\beta$ of the cavity from the initial beam transport analysis was found to be 0.5. The beam transport simulation was repeated using the on-axis field distribution from the electromagnetic field solver used to design the cavity, as described in Sec. IIIA. Figure 4 shows the beam evolution as it travels downstream of the cathode. The electron bunches exit the thermionic gun with a kinetic energy of 100 keV (Fig. 4a). The solenoid then focuses the bunch into the accelerating cavity (Fig. 4b). The phase and gradient of the cavity are set to -9° and

5.6 MV/m (13.6 MV/m peak electric field on-axis), respectively, such that the bunch leaves with an average kinetic energy of 1 MeV. To achieve 100% transmission of the bunch through the cavity, the required phase results in some bunching and transverse focusing to the electron beam (Fig. 4c). This process unavoidably increases the energy spread of the bunch (Fig. 4d). Details of the energy distribution after the cavity, at a distance of 2.4 m from the cathode are shown in Figure 5. While there is an energy tail to the distribution, 85% of the bunch is within +/- 100 keV of the target energy. The overall envelope of the beam is less than 4 cm in radius before the cavity and less than 5 cm in radius after the cavity. Given that the radiuses of the cavity beam tubes are 5.5 cm and 9 cm (see Sect. III.A) , no electrons would be lost on the beam chambers' walls according to the simulation.

    These simulations show in this simplified case that the resulting beam is suitable for industrial purposes. Further development would require designing the specific electrode geometry in the thermionic gun and specify the RF drive to the cathode grid to provide the longitudinal bunch profile at the cathode. Further improvements to the energy spread, if required, could be made with the addition of a normal conducting buncher cavity before the SC accelerating cryomodule. These cavities typically operate with a few kW of power as they are operated such that there is no net energy gain of the electron bunch. These options can all be easily explored through further simulations in the future.

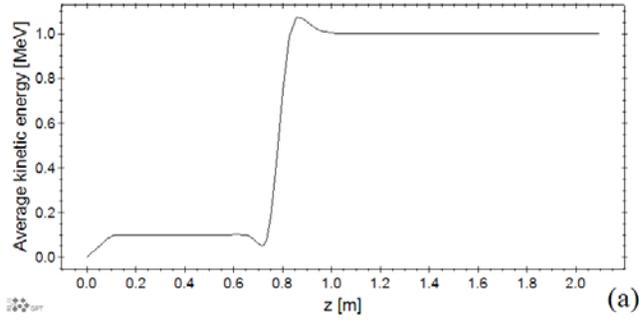
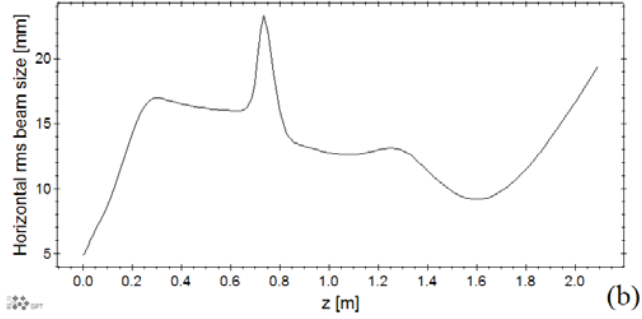
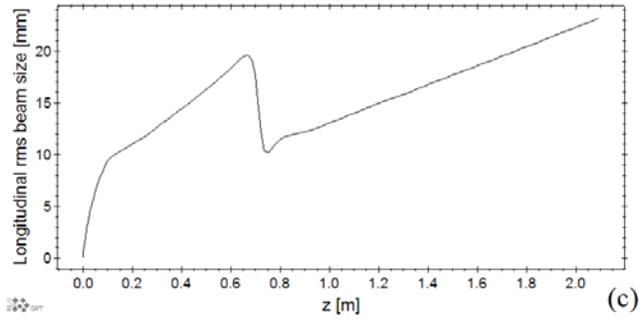
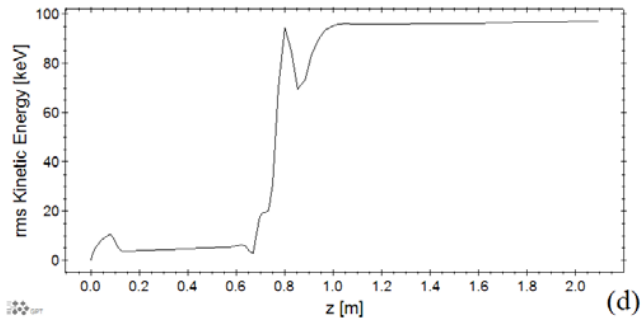

FIG. 4. Beam evolution through the accelerator: average kinetic energy (a), transverse rms bunch size (b), longitudinal rms bunch size (c) and rms energy spread (d).

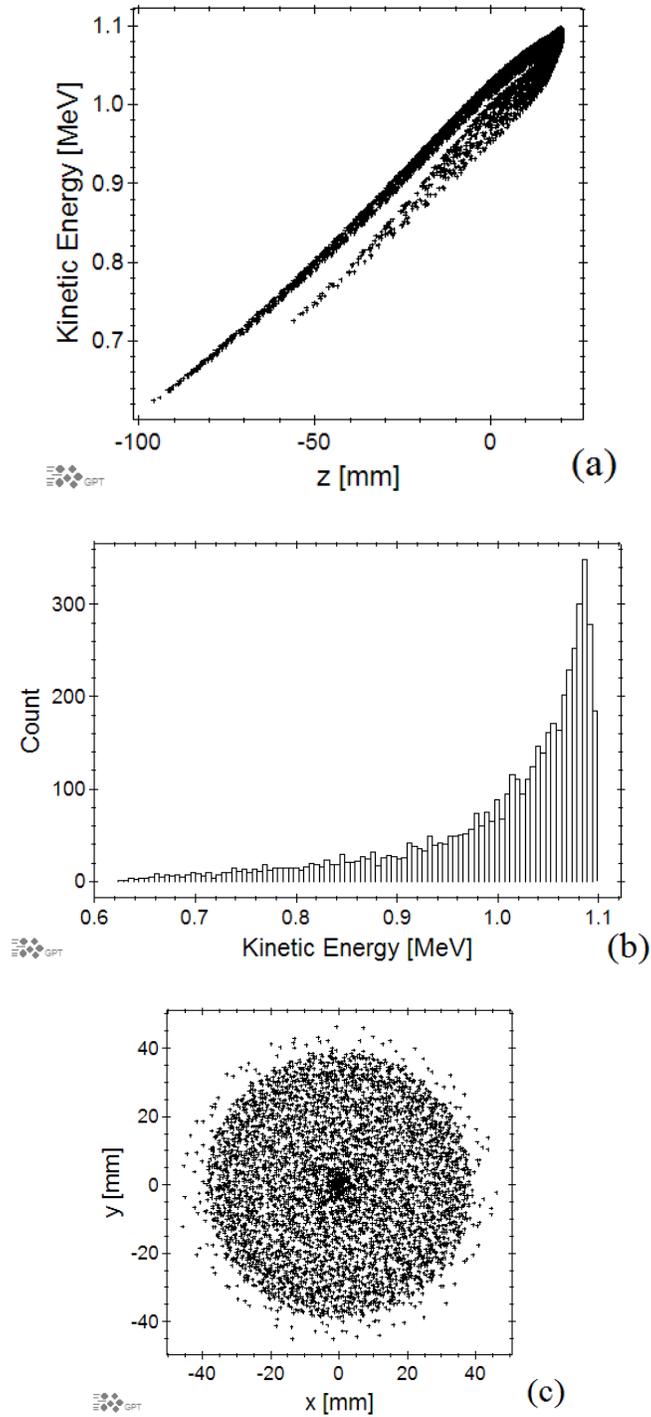

FIG. 5. Longitudinal phase space at 2.4 m from the cathode (a) and associated histogram (b). The transverse configuration space after the cavity is shown in (c).

## III. SRF CRYOMODULE DESIGN

Standard SRF cryomodules used in particle accelerator for physics or material science research rely on liquid He refrigerators to supply large volumes of liquid He to cool the SRF cavities, typically made of bulk Nb, to ~4.3 K or ~2 K. The implementation of the SRF technology in a compact, low-cost, high reliability industrial accelerator requires an alternative method to cool the SRF cavity well below the transition temperature. Gifford-McMahon type cryocoolers have proven to be a reliable, low-cost option to cool superconducting magnets used in magnetic resonance imaging machines typically installed in hospitals. Advances in cryocooler technologies result in units with increasing cooling power at 4 K. One such unit, SRDK-415D from Sumitomo Cryogenics of America, provides a cooling power up to 1.5 W at 4.2 K.

Given the limited available cooling power, the SRF cavity material should have lower surface resistance than that of bulk high-purity Nb at 4 K. Recent R&D efforts on $Nb_3Sn$ for SRF cavity application have showed the possibility of achieving quality factors of ~$1\times10^{10}$ at 4.3 K and accelerating gradient, $E_{acc}$, of up to ~16 MV/m in 1.3 GHz single-cell cavities [14]. The possibility to achieve such performance makes $Nb_3Sn$ the ideal superconductor for this application. $Nb_3Sn$ is formed as a thin film by thermal diffusion of tin onto the surface of a bulk Nb cavity at high temperature, greater than ~1000 °C.

Cooling of the cavity with cryocoolers occurs by conduction along the cavity walls, therefore it is important to have a high-thermal conductivity material, such as copper, deposited onto the outer surface of the Nb cavity with the thin film $Nb_3Sn$ on the inner surface. Deposition of copper onto the outer cavity surface could be done, for example, by electrodeposition, vacuum plasma spray or gas dynamic cold spray.

## A. Cavity design

The following considerations determined the choice of the cavity frequency: for a fixed energy gain of ~1 MeV, the required accelerating gradient increases linearly with the cavity frequency, $f$. The cavity surface area decreases as $1/f^2$, therefore the heat flux due to rf losses, $q$, scales as:

$$q = \frac{\kappa}{d}\Delta T \propto R_s f^4, \qquad (1)$$

where $\kappa$ is the thermal conductivity of the cavity wall, $R_s$ is the surface resistance of the superconductor, $d$ is the distance between the point of the cavity surface furthest away from the cryocooler point-of-contact with the cavity and $\Delta T$ is the temperature difference between that point and the cryocooler temperature (4.3 K). $d$ increases as $1/f$, therefore $\Delta T$ is proportional to:

$$\Delta T \propto \frac{R_s}{\kappa} f^3 \qquad (2)$$

$R_s$ is the sum of two components, one ($R_{BCS}$) related to the quasi-particle density and proportional to $f^2$, as predicted by the Bardeen-Cooper-Schrieffer theory of superconductivity, and one component ($R_{res}$) related to the presence of defect, normal conducting inclusions, trapped magnetic flux, etc. whose frequency dependence is not well studied. From Eq. (2), it is evident that low-frequency cavities are favored in order to achieve the most uniform temperature on the cavity inner surface. However, lower frequency implies larger cavity size, impacting the size of the cryomodule, and larger surface area, increasing the probability of local defects limiting the cavity rf performance. As a compromise between these competing arguments, a frequency of 750 MHz was chosen.

As shown in Sec. II, in order to properly accelerate the non-relativistic beam out of the gun, the geometric β of the cavity needs to be 0.5. Given the complexity of the multi-layered cavity material and the preference for the largest possible iris diameter to minimize any possibility of beam loss, an elliptical-shape type cavity was preferred over a spoke-type cavity.

The cavity geometry was optimized using Superfish v. 7.17 [15] with the user interface BuildCavity [16, 17] which provides a parameterization of the geometry and automatic tuning to the desired frequency. A schematic layout of the cavity shape is shown in Fig. 6. The shape was optimized to lower the peak surface electric and magnetic fields, $E_p$ and $B_p$, respectively, while keeping a large iris diameter, adequate mechanical stability and ease of surface processing, which limited the wall angle to 7°. The diameter of the beam tube was enlarged on one side of the cavity to allow for propagation of HOMs. The radius of the transition between the iris diameter and beam tube diameter was chosen as to avoid any local minimum of the electric field, avoiding possible multipacting in this region [18]. A summary of the electromagnetic parameters is given in Table I. The two coaxial FPCs are positioned along the larger beam tube and their axis, 180° apart, at 65 mm from the cavity iris. The $Q_{ext}$ of the FPCs is given by:

$$Q_{ext,FPC} = \frac{V_{acc}}{R/Q \, I \cos\varphi} = 4.8 \times 10^4. \tag{3}$$

The tip of the FPCs would follow the contour of the beam tubes ("pringle" shape) to obtain a low $Q_{ext}$ while minimizing wakefield effects [19]. Acceleration to ~1 MeV corresponds to operating the cavity at an accelerating field of 5.6 MV/m, which would correspond to rather modest peak surface fields: $E_p$ = 19.3 MV/m and $B_p$ = 40.5 mT. Operation at these peak surface fields should reduce the probability of the cavity from being limited by field emission or low $Q_0$. Multipacting analysis of the cavity geometry was done using the code FishPact [20]. Stable two-point electron trajectories close to the equator with impact energy of ~25-45 eV were found for $E_{acc}$-values in the range 7-9.5 MV/m. The impact energy of those electrons is close to the threshold corresponding to a secondary yield greater than unity in $Nb_3Sn$ surfaces [21, 22] and could result in multipacting. However, the predicted possible multipacting barrier is above the operating gradient, and it can be significantly suppressed by the detailed real geometry at the equator, where a weld is located [23].

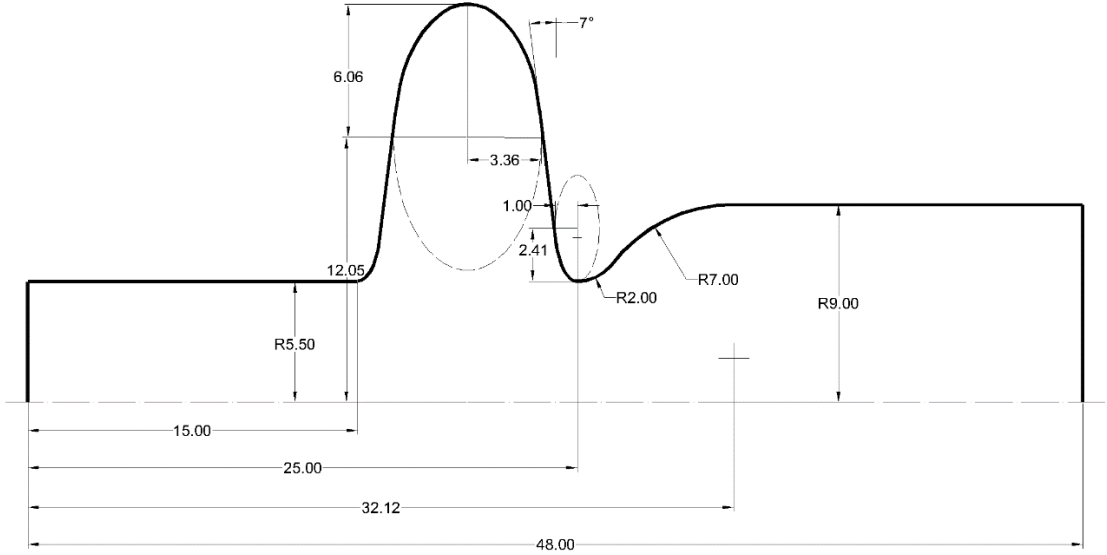

FIG. 6. Geometry of the β=0.5 SRF cavity. Dimensions are in centimeters.

TABLE I. Summary of the electromagnetic parameters of the single-cell cavity

| | |
|---|---|
| $f$ (MHz) | 749.973 |
| $E_p/E_{acc}$ | 3.44 |
| $B_p/E_{acc}$ [mT/(MV/m)] | 7.23 |
| $G$ (Ω) | 155.4 |
| $R/Q$ (Ω) | 21.3 |

**B. HOM analysis**

One objective of the cavity design was to damp HOMs to an adequately safe level to avoid potential transverse single pass beam breakup (BBU) instabilities. This has led to the consideration of enlarging the beam tube on one side (ID = 180 mm) of the cavity as stated above. The larger beam tube has a first cutoff frequency of 0.976 GHz ($TE_{11}$) and 1.275 GHz ($TM_{01}$) for dipole and monopole modes, respectively. This allows the first dipole mode pair ($TM_{110}$) as well as the first monopole mode to propagate out of the cavity towards the BLA. The smaller beam tube (ID = 110 mm) has a first cutoff frequency 1.597 GHz ($TE_{11}$) and 2.086 GHz ($TM_{01}$). During the design optimization, a beam tube ID = 116 mm had been conceived utilizing two or four flutes to lower the cutoff frequency, similar to the CESR-B cavity developed at Cornell. However, this option was abandoned to simplify the engineering design and the complexity of the flange connection from the cavity to the beam pipe.

The HOM damping also takes into account that RF fields may propagate out of the two coaxial FPCs. The broadband coupling impedance for dipole and monopole HOMs as excited by a beam of finite bunch length has been numerically calculated utilizing the CST STUDIO SUITE® (CST) wakefield solver code [24]. Since a wakefield computation needs to be aborted after a finite time/length, an extrapolation of the full impedance spectrum is done to infinity [25], which improves the peak impedance resolution. The wakefield computations assume a Gaussian-shaped bunch with finite length ($\sigma_{rms}$). The impedance amplitude spectra have been normalized by the bunch spectrum, which leads to an overestimation of the impedance at the far frequency end as the bunch spectrum tails off rapidly. Therefore only a fraction of the impedance spectrum below the far frequency end of the spectrum is taken into account generally[1]. For the simulations $\sigma_{rms}$ = 30 mm has been assumed to limit the CPU time to reasonable values, while covering all high impedance HOMs for the BBU analysis.

In parallel to wakefield calculations, complex (with lossy materials) CST Eigenmode calculations [26] have been performed to verify the resolution of the most crucial impedances.

---

[1] A good estimation of the upper frequency limit is f = $2c/(\pi \cdot \sigma_{rms})$, where the bunch spectrum rolls off to about 1/3 ‰ of its value at zero frequency.

The Eigenmode solver allows utilizing tetrahedral meshes for a more accurate discretization of the RF volumes than achievable with the hexahedral meshes mandatory for the wakefield computations. The Eigenmode solver provides both the $R/Q(\beta)$ and external $Q$ value of each mode from which the HOM impedance is calculated. For the longitudinal, $R_l$, and dipole impedance, $R_{tr}$, the following definitions have been used:

$$R_l = \frac{R(r=0)}{Q} \cdot Q_L \; [\Omega] \tag{4}$$

$$R_{tr} = \frac{R(r)}{Q} \cdot Q_L \frac{1}{k \cdot r^2} \; [\Omega/m] \tag{5}$$

Herein $k = \omega/c$ is the wave number, $r$ the radial offset from the cavity axis and $Q_L$ is the loaded $Q$. The circuit definition $R(r)/Q = V(r)^2/(2Q \cdot P_{avg}) = V(r)^2/(2\omega U)$ is chosen throughout, wherein $V(r)$ represents the transit-time corrected voltage along the cavity and $P_{avg}$ and $U$ the average power and cavity stored energy, respectively. The loaded $Q$ is equal to the combined $Q_{ext}$ resulting from absorption at all outgoing ports (large beam tube, small tube, two FPC ports beyond RF window) assuming negligible RF surface losses in the cavity.

Note that the wakefield computations require a speed of light beam excitation ($\beta = 1$) to provide meaningful results though the beam is yet not ultra-relativistic. For the Eigenmode simulations the β-value is fixed but variable. A comparison of the impedance values resulting from both methods revealed that the discrepancy is not significant. The dipole impedance spectrum up to 4.5 GHz is shown in Fig. 7. Hereby two wakefield calculations were performed with the beam exciting the wakefield with a radial offset either in horizontal (H) or vertical (V) direction to study the damping of the differently polarized dipole mode pairs. Any difference implies a preferential damping due to the FPCs, which are pointing in the horizontal direction. For instance, the first horizontally polarized dipole mode ($TM_{110}$) is better damped than its vertical polarization.

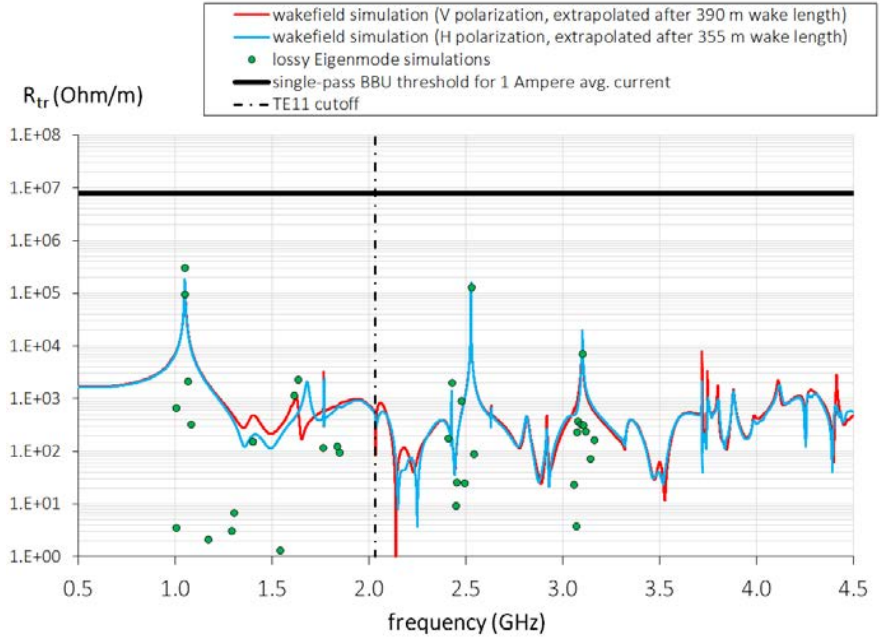

FIG. 7. Cavity dipole impedance spectra (solid lines = wakefield calculations with beam traversing off axis with a radial offset *r* either in horizontal vertical direction, green dots = complex Eigenmode computations, either polarization). The horizontal black line is the single pass BBU impedance threshold impedance for a 1 A average beam current.

Since all HOMs are allowed to propagate through the beam tubes, one has to conceive that the *R*/*Q*-values may depend significantly on the integration path (beam tube length) and that the net interaction of the beam with the traveling fields may result in residual *R*/*Q* or impedance values smaller than calculated with the rather finite beam tube lengths as modelled.

For an estimate of the single pass BBU threshold current ($I_{threshold}$) for a specific HOM, one may utilize the following equation for standing-wave regenerative oscillations [27]:

$$I_{threshold} \approx \frac{(\frac{\pi}{2 \cdot L_{act}})^2 \cdot \lambda_{HOM} \cdot V_{beam}}{\frac{R(r)}{Q} \cdot \frac{1}{k^2 \cdot r^2} \cdot Q_L \cdot \frac{1}{L_{act}}} = \frac{\pi^3 \cdot V_{beam}}{2 \cdot L_{act} \cdot R_{tr}} \qquad (6)$$

$L_{act}$ denotes the active cavity length (0.1 m from iris to iris), $\lambda_{HOM}$ the wavelength at the HOM frequency and $V_{beam}$ the beam energy. Solving for the transverse impedance, we then can calculate the BBU threshold impedance as plotted in Fig. 7 (horizontal black line) for an average beam current of 1 A. As a conservative estimate we assumed a beam energy of 100 keV as delivered from the injector. All crucial HOM impedances therefore have a margin of more than one order of magnitude to the BBU impedance threshold.

Figure 8 shows the monopole spectrum up to 4.5 GHz. The monopole modes are dominating the total HOM power losses that can be deposited in the SRF cavity. These power losses need to be accounted for to provide adequate cooling of coupler components, particularly to handle the power dissipated in the room temperature BLA, which will drive its design.

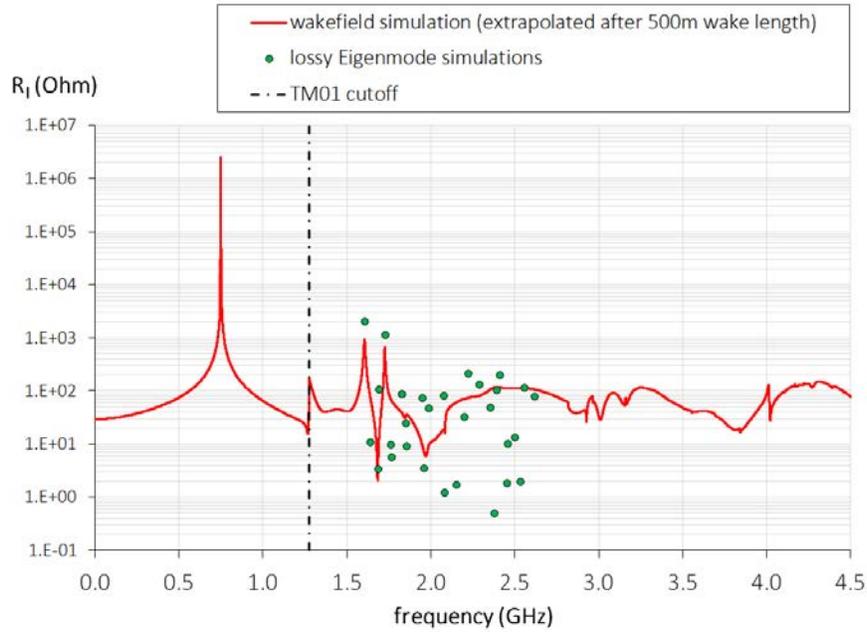

FIG. 8: Cavity monopole impedance spectra (solid line = wakefield calculation with beam traversing along axis, green dots = complex Eigenmode computations, either mode polarization). Note that the accelerating mode is not resolved in the wakefield simulation. The external $Q$ is dominated by the FPC couplers.

Though the simulations have been limited to a bunch length of $\sigma_{rms} = 30$ mm due to the reason stated above, an estimate of the total HOM power up to a much higher frequency is required given that the bunch length is $\sigma_{rms} \sim 10$ mm at the cavity entrance. The power deposited in the higher frequency regime (up to 16.5 GHz) had been estimated assuming that there is no high-Q/high-impedance mode existing to be resonantly excited with high power levels. This is justified by the fact that the HOMs propagating out of the cavity at higher frequencies generally exhibit very low Q-values. Furthermore, the beam spectral lines in CW operation are spaced apart only every 750 MHz. As a (rather worst case) estimate for the unknown mode impedances in the spectral regime of 6.75-16.5 GHz it was assumed that the normalized impedance of the HOMs is constant rather than rolling further off and equal to that calculated at 6 GHz. The product of the real impedance and the peak current yields the power deposited at each spectral line. The rationale is illustrated in Fig. 9, where the real part of the normalized impedance spectrum is plotted as calculated with CST together with the beam current spectral lines for an average current of 1 A taking into account the roll-off of a 10 mm rms bunch length as illustrated. The total power in this way sums up to ~213 W. Only about 10% of the power is contributed by the spectral lines within 6.75-16.5 GHz. The BLA might experience a significant fraction of the total power deposition. The power for the first two spectral lines (1.5 GHz and 2.25 GHz) is ~40 W and 75 W, respectively, and safely away from the two monopole HOMs with the highest impedance at the low frequency end (compare Fig. 8), which avoids resonant excitation.

In conclusion, the power deposition in the BLA is relatively small. For comparison, the CESR and KEK-B BLAs, consisting of water-cooled lossy ferrite material, have demonstrated to withstand several kW of HOM power in operation (successful tests were conducted up to 10.8 kW already in 1999 [10]).

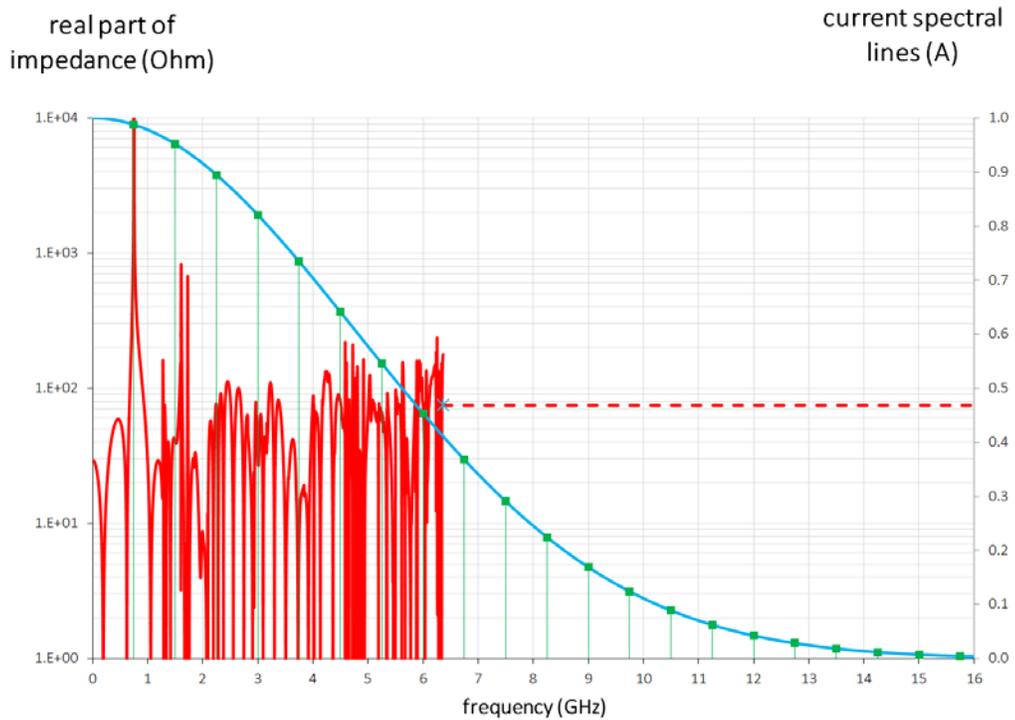

FIG. 9. Real impedance of the monopole modes (red line) and beam current spectral lines (green vertical lines) for an average current of 1 Ampere. See text for detailed explanation.

## C. Thermal analysis

The solid model of the cryomodule used for the thermal analysis with the finite element code ANSYS [28] is shown in Fig. 10. The model includes the $Nb_3Sn/Nb/Cu$ cavity with Ti45Nb flanges, stainless steel bellows, and Cu thermal shield attached to the first stage of the cryocoolers. The bellows shown in the figure provides for thermal growth differences as well as thermal isolation. The different colors in the Fig. 10 represent different materials. The thermal intercepts and power coupler pringle as well as the inner conductor are made of copper.

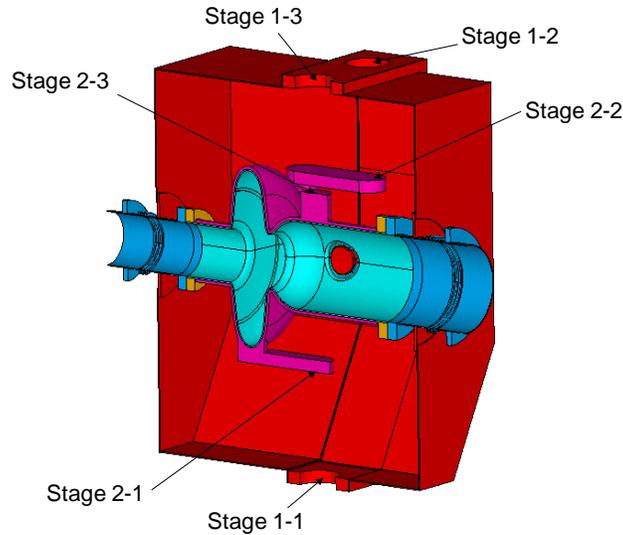

FIG. 10. 3D model used for the thermal analysis. Each color represents a different material: the lighter blue indicates $Nb_3Sn$, the darker blue stainless steel, the orange color represents Ti45Nb, the red is the copper first stage intercept and the fuchsia color represents the copper plating on niobium. The nomenclature "Stage 1-3" means first stage of cryocooler number 3. The fourth cryocooler is symmetric to cryocooler number 2.

The temperature-dependent thermal conductivity of the model materials is shown in Fig. 11 [29]. The temperature-dependent surface resistance of $Nb_3Sn$ at 750 MHz, shown in Fig. 12, was determined using a computer code [30] which calculates the surface resistance according to the Bardeen-Cooper-Schrieffer theory of superconductivity, to which a temperature-independent residual resistance of 10 n$\Omega$ was added to account for extrinsic effects such as trapped magnetic flux and defects. The material constants used for the calculation of the surface resistance are: a critical temperature of 18 K, a London penetration depth of 160 nm, a coherence length of 4.7 nm and a mean free path of 3 nm. The heat capacity of the envisioned cryocooler SRDK-415D is shown in Fig. 13 as a function of the temperature of the first and second stage.

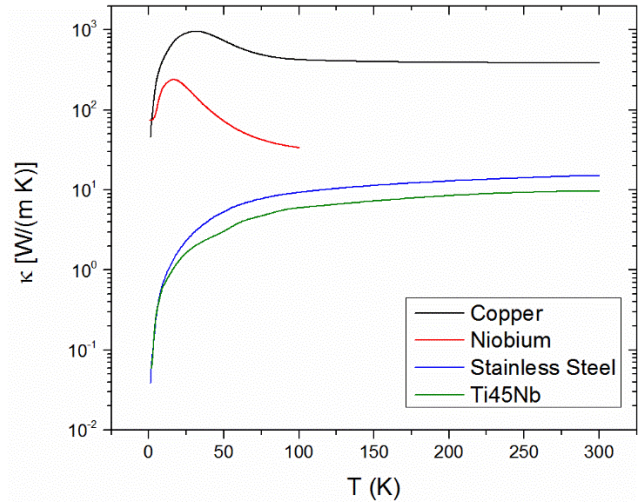

FIG. 11. Temperature dependent thermal conductivity of the materials used in the thermal analysis.

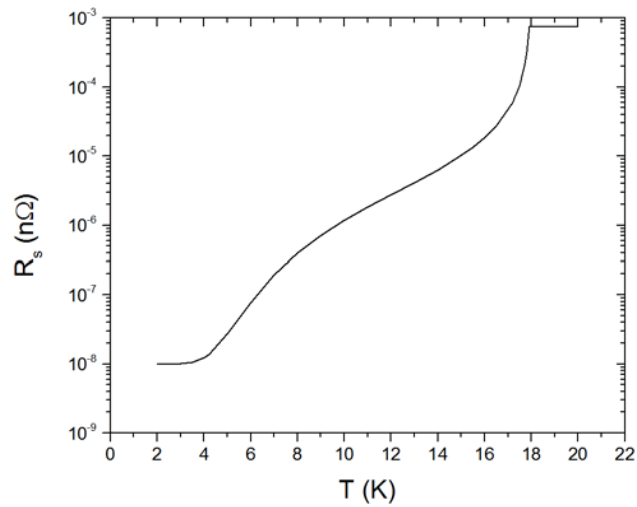

FIG. 12. Temperature dependent surface resistance of $Nb_3Sn$ used in the thermal analysis.

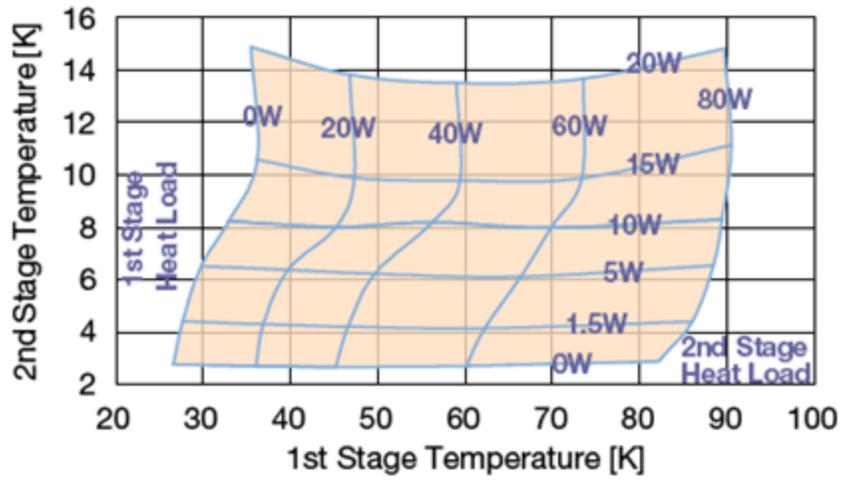

FIG. 13. Cooling power as a function of temperature for the first and second stage of a commercial cryocooler (SRDK-415D, Sumitomo Cryogenics of America) [31].

The different stages of the cryocoolers are represented by surfaces which have their temperature set for the subsequent iteration based on the heat loads of both stages from the prior iteration. The ambient temperature of 300 K was set at the power coupler flange that connects to the cryostat and at the beam pipe flange that connects to the cryostat. The thermal radiation between inner and outer conductors of the FPCs is included in the analysis. The proportion of radiative heat that leaves a surface and strikes another one ("view factor") was calculated for the surfaces of the FPCs with a finite element code. An emissivity value of 0.02 is assumed for Cu, whereas it is 0.2 for $Nb_3Sn$. The thermal radiation between regions with multilayer insulation, such as between the cryomodule outer walls and the thermal shields and between the thermal shields and the inner components, is not included in the analysis but it is expected to have a minor effect. An RF power of 600 kW is assumed to flow in each of the FPCs, and the cavity surface fields correspond to the nominal accelerating gradient of 5.6 MV/m. The FPCs outer conductors are made of stainless steel with a 15 μm thick Cu coating.

A converged solution, shown in Fig. 14, was found with a minimum of four cryocoolers, three having their second stage attached to the cavity and one having the second stage on the beam tube to intercept the heat from the two FPCs. Large thermal gradients occur between the room temperature flanges and the thermal shields through the stainless steel bellows. The maximum temperature, 302 K, is at the power coupler pringle that is heated by RF and cooled by room temperature coolant. The thermal shields with 2.29 mm (0.090") thick walls provide enough cooling to maintain acceptable temperatures within the beam pipe to the second stage cryocooler connection. The first stage temperatures are near 40 K with heat loads near 20 W. Details of the cavity temperature distribution and heat loads to the second stages are shown in Fig. 15, indicating a stable cavity temperature of ~6 K. A summary of the heat loads and the temperatures of the first and second stages of the cryocoolers is given in Table II, whereas a summary of the cryomodule heat loads is given in Table III.

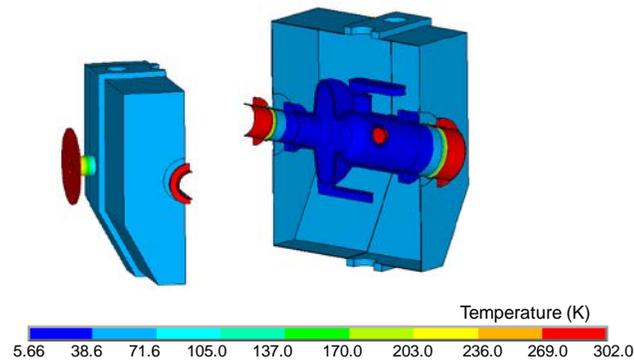

FIG. 14. Temperature distribution and boundary heat loads for the single-cavity operating at a gradient of 5.6 MV/m and with 600 kW RF power into each coaxial FPC.

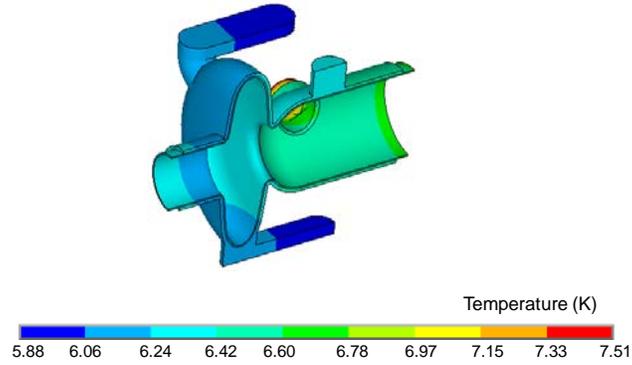

FIG. 15. Temperature distribution along the cavity surface.

TABLE II. Summary of the heat loads and temperatures of the cryocoolers

|  | Cryocooler 1 | Cryocooler 2 or 4 | Cryocooler 3 |
|---|---|---|---|
| Stage 1 heat load (W) | 22.6 | 20.1 | 19 |
| Stage 1 temperature (K) | 40 | 39 | 39 |
| Stage 2 heat load (W) | 4 | 3.94 | 4.65 |
| Stage 2 temperature (K) | 5.7 | 5.7 | 6.1 |

TABLE III. Summary of the cryomodule heat loads

| | |
|---|---|
| Total heat load into cryocoolers' stage 1 | 81.2 W |
| Total heat load into cryocoolers' stage 2 | 16.5 W |
| Cavity dynamic heat load | 2.8 W |
| Static heat load from cavity end-groups | 9.6 W |
| Two FPCs dynamic heat load | 32.6 W |
| Two FPCs static heat load | 52.6 W |
| Two FPCs center conductor heat load | 268 W |
| Radiant heat load from FPCs center conductor | 0.7 W |

The effects of applying a thin film high-temperature superconductor such as YBCO to the inner surface of the FPCs outer conductors from the cavity to the location where the temperature approaches 90 K was evaluated. Figure 16 shows the temperature-dependent surface resistance of YBCO, $f^2$-scaled from data at 8 GHz [32], with the surface resistance of copper at 90 K. The model results suggest that only three cryocoolers would be needed if a YBCO coating could be reliably applied to the copper. The second stages of two cryocoolers would intercept the cavity and the second stage of the third cryocooler would intercept the FPCs.

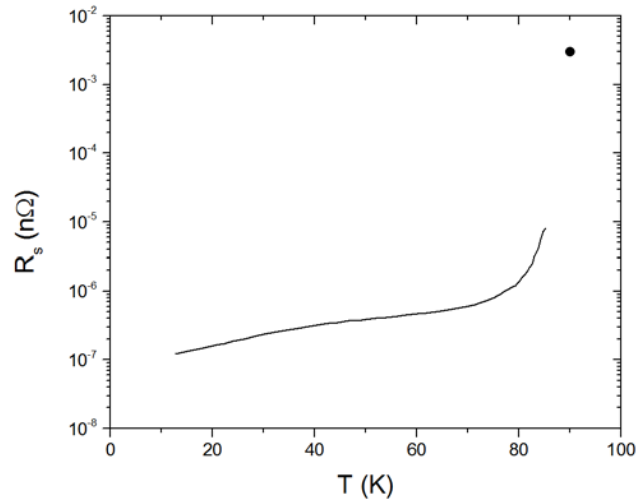

FIG. 16. Temperature dependence of the surface resistance of YBCO films scaled to 750 MHz from values measured at 8 GHz [32]. The solid symbol is the surface resistance of Cu at 90 K.

## D. Cryomodule design and assembly

In this section we will present the cryomodule design and the details of how we envisage fabrication and assembly of the cavity and cryomodule. Many different configurations were considered in the design/analysis process. We started with concepts employing three waveguide feeds to the cavity. This configuration offered the attractive feature that the waveguides can act both as the RF power input couplers and the HOM out-couplers, which conserves real estate in the hermetic assembly. However, such configuration would have required six cryocoolers to achieve a stable thermal condition, due to the combination of dynamic RF losses on the waveguides during operation and the large thermal conduction cross section of the waveguides.

The solution we finally chose for concept development employs dual coaxial power couplers of a type used at these power levels before and a beamline HOM absorber of the Cornell CESR-B type, also an established design with many in service. The coaxial power couplers offer a small conduction cross section and allow us to remove the bulk of the dynamic RF losses with water cooling of the inner conductor, avoiding further cryogenic load. This system exhibits thermal stability with margin utilizing four cryocoolers. Furthermore, we attempted to keep the cryocoolers in the vertical orientation to the greatest extent possible as this is the most efficient condition. Our design keeps three in the vertical and one in the anti-vertical orientation. The following paragraphs describe the design in detail starting with the SRF cavity and building out to the full assembly.

*1. Cavity fabrication*

The cavity is a single-cell, 750 MHz elliptical cavity conventionally fabricated from 3 mm thick high residual resistivity ratio niobium sheets. The beamline flanges and power coupler flanges are Ti45Nb electron-beam welded to the niobium tubes, with the same sealing design as that developed at DESY [33]. In addition, there will be a small port in the upstream beam pipe (not shown in the present layouts) for an RF pickup probe. The output beam tube features an expanded diameter just past the cavity iris to facilitate transmission of the HOMs.

Once the cavity is fabricated it will be chemically processed with standard methods, such as buffered chemical polishing or electropolishing, and the $Nb_3Sn$ film will be grown on the inner surface by the conventional vapor diffusion method, which involves annealing the cavity at high temperature (~ 1200 °C) in the presence of a Sn vapor [14].  When the cavity and coating has passed acceptance tests, the cavity will be sealed and put through the electroforming process to deposit approximately 6 mm thick of high purity copper on the surface.  This process is being investigated on niobium coupons at Jefferson Lab and resulted in a RRR of the copper layer of ~300 [34].  Though the technology of the electroforming is well understood, the electroforming procedure will need some development to allow us to build up the areas that will interface to the cryocoolers.  These areas will need substantially more copper deposited or perhaps a combination of additional electroform and solid copper pieces "grown-in" to the overall deposit.  There is no technical barrier to achieving this, however it is a time consuming process that may involve iterations of electroforming and machining.  Figure 17 shows the cavity after electroforming and post electroform machining.

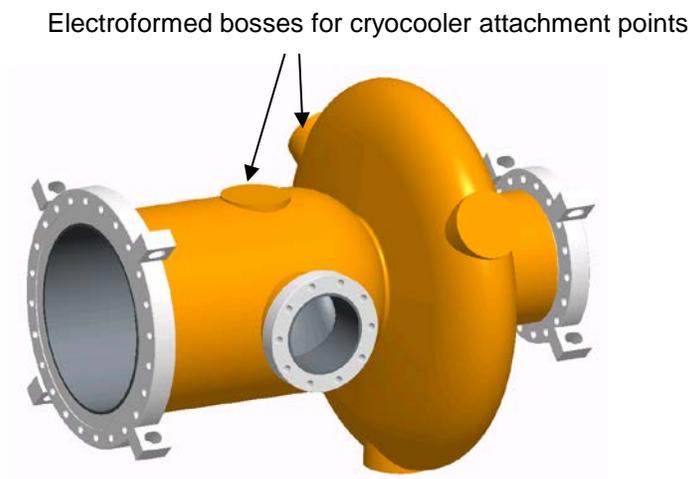

FIG. 17. 3D model of the cavity after electroforming a 6 mm thick Cu layer on the outer surface.

*2. Hermetic string*

The hermetic string in this design, shown in Fig. 18, comprises the copper coated SRF cavity, thermal isolation sections of the beam tubes upstream and downstream for the transition from 4 K to room temperature, a beamline HOM damper of the Cornell CESR-B type on the downstream side and all metal RF sealed isolation valves on each end. Two FPCs capable of 600 kW CW are also part of the hermetic assembly. These FPC's are very similar in design to those built by Communication & Power Industries (CPI) and operated on the Brookhaven/AES Superconducting RF electron gun [35]. The difference lies in the cooling of the vacuum side outer conductor. In the design for BNL this region was cooled by a 5 K gas stream in a spiral channel. In this design we incorporate a copper jacket running part way to the 50 K intercept that conductively cools this region. All of these heat loads have been considered and included in the analysis presented in this report. A coupler of the type built for the BNL/AES SRF gun is shown in Fig. 19.

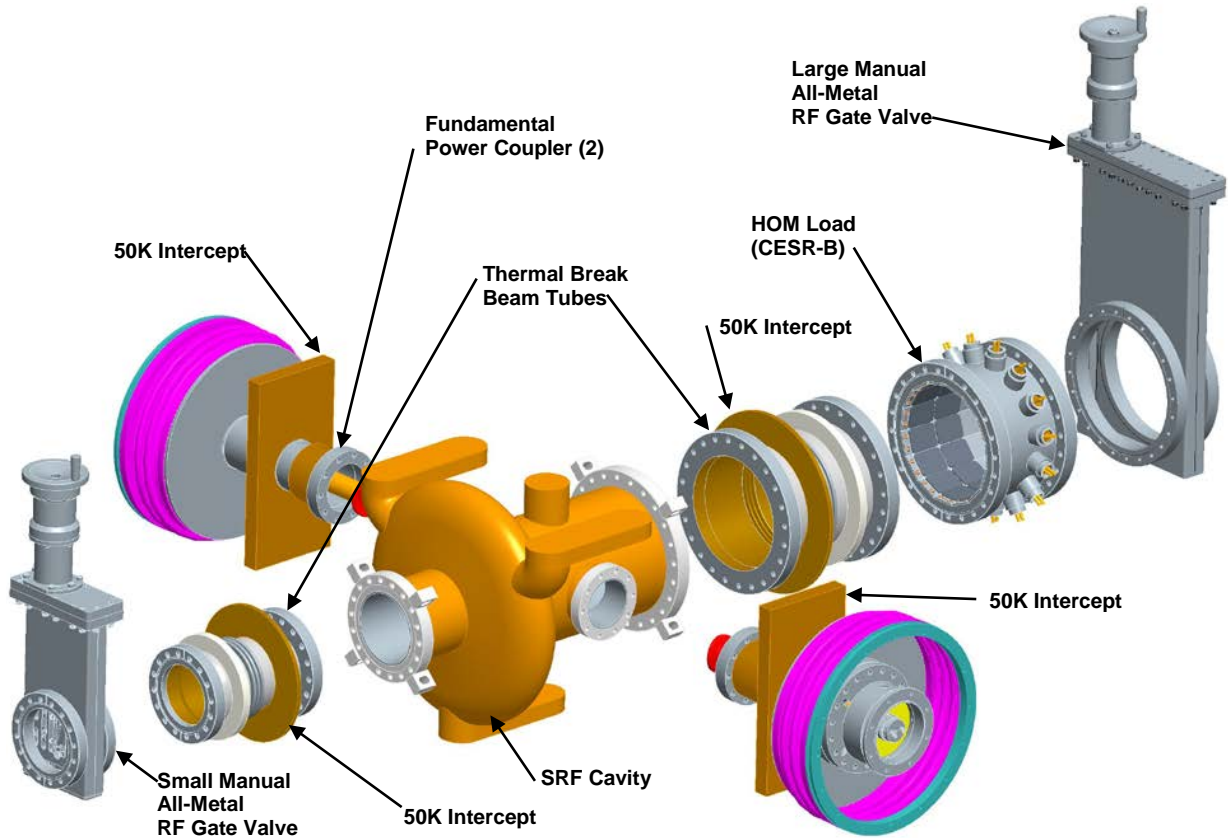

FIG. 18. 3D model of the components comprising the hermetic string to be assembled in the clean room.

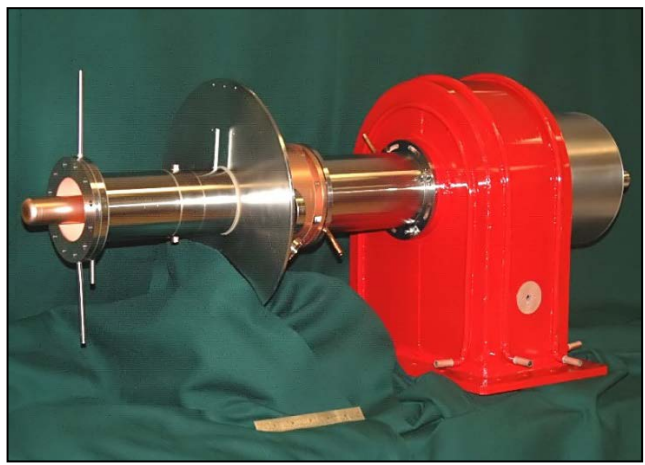

FIG. 19. MW-class, 748 MHz CW coaxial FPC manufactured by CPI (model VWP1185/1186) [36].

*3. Cryomodule assembly*

The first assembly to be made outside the clean room on the hermetically sealed string is the core structure of the 50 K thermal shield. This structure will be made from oxygen-free electronic (OFE) copper plates with high thermal conductivity strain relief breaks in several locations to prevent mechanical overloading of the cryocoolers during cooldown. This core structure is where the first stage connections from the cryocoolers will be attached, while the second stages will be attached at the corresponding locations on the cavity. Copper thermal straps such as the model P5-502 from Technology Applications, Inc. [37] are one of the possible type of strain isolators envisioned for both the 50 K shield structure and the 4 K cavity connections.

The core assembly is now installed in the cryomodule vacuum vessel as illustrated in Fig. 20. The cold mass will be supported from the cavity end flanges by stainless steel tension rods. The isolation valves will ultimately be supported by the vacuum vessel end walls. Once in the vacuum vessel the string will be aligned and assembly will continue.

At this point the cryocoolers will be installed and connected to the cavity 4 K interfaces, the shield 50 K interfaces, and the vacuum vessel interfaces. Installation of the cryocoolers will be followed by installing the FPC Top-Hat assemblies that complete the vacuum boundary of the FPCs. Also at this time the instrumentation and wiring will be installed followed by inner MLI blankets. The next step in assembly is the inner magnetic shield. We expect this to comprise approximately twelve pieces that can be assembled over and around the cavity, cryocoolers, and supports. Once installed, additional MLI blankets will be applied. Figure 21 shows the cryomodule with the inner magnetic shield installed (blue color).

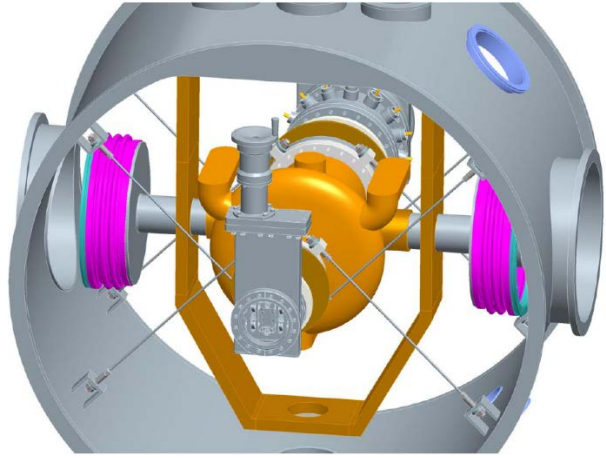

FIG. 20. Hermetic string inserted into the vacuum vessel.

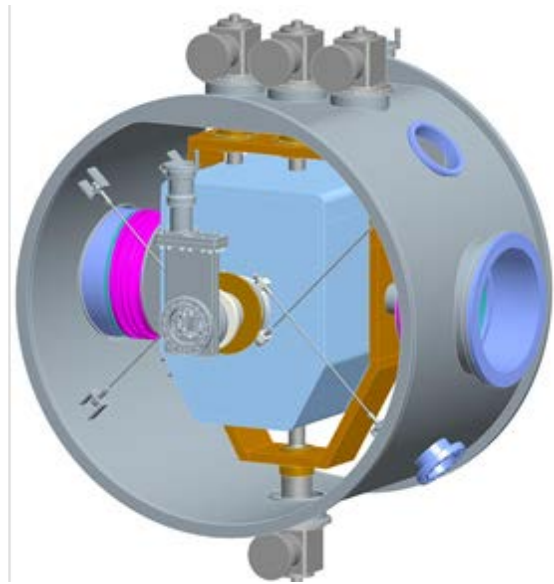

FIG. 21. Cryomodule with cryocoolers, FPC top-hats and inner magnetic shield installed.

The next step after the inner magnetic shield is the installation of the balance of the thermal shield. As with the magnetic shield, this will be a multi-piece assembly comprising approximately ten pieces that will be connected to the center 50 K assembly. Connections will also be made at the beam tube thermal transitions. The detailed design of the thermal shield will require careful consideration to preserve the strain isolation for the cryocoolers. Once again after installation, additional MLI blankets will be applied.

The final step of the internal assembly of the cryomodule is the installation of the outer magnetic shield. This assembly will comprise approximately fourteen pieces that will again assemble over and around the core assembly, cryocoolers, and support structure. It may indeed prove desirable to assemble the upper and lower tray pieces prior to installation of the cavity string into the vacuum vessel but this detail will be developed as design progresses. The use of two layers of magnetic shielding is common practice in the design of SRF cryomodules to shield the Earth's field to a residual magnetic field of less than 10 mG inside the the inner shield [38, 39, 40]. Figure 22 shows the cryomodule after installation of the thermal shield and of the second magnetic shield.

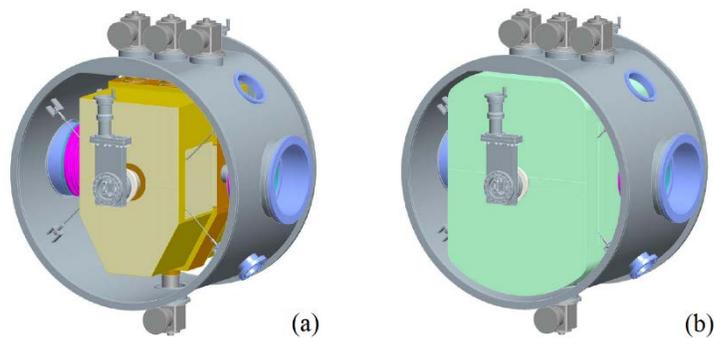

FIG. 22. Cryomodule after installation of thermal shield (a) and outer magnetic shield (b).

The final step in the completion of the cryomodule is the installation of the split end-plates on the cylindrical portion of the vacuum vessel. The two halves of each end-plates would be welded together and to a ring which is part of the beamline assembly. The module can then be transported to the installation site. Once in its intended location, the final stage of assembly is the addition of the air-side components of the FPCs, which transition the coaxial coupler units to the WR1150 waveguide transmission system. This will be followed by completion of the hookup of RF, water, air, instrumentation, and vacuum systems. Figure 23 presents views of the completed cryomodule and the overall dimensions are listed in Table IV.

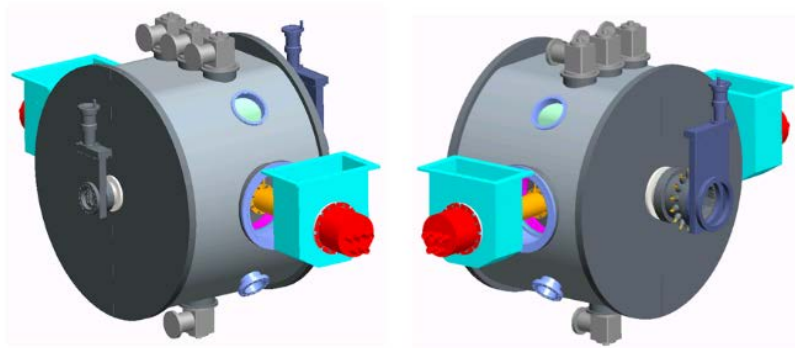

FIG. 23. 3D views of the fully assembled cryomodule.

TABLE IV. Overall cryomodule dimensions

| | |
|---|---|
| Length | 1.26 m (49.5") |
| Width (across FPCs) | 2.52 m (99.2") |
| Vacuum vessel flange diameter | 1.17 m (46") |
| Height (across cryocoolers) | 1.48 m (58.4") |

## IV. ENGINEERING AND COST ANALYSIS

In this Section, we will discuss engineering aspects related to the operation of the 1 MW SRF accelerator, such as the RF power source, the low-level RF and beam diagnostic and radiation shielding. An estimate of the efficiency, the cost analysis and a conceptual layout for a commercial facility are also discussed in this Section.

### A. RF Power Amplifiers

Although solid state has become the predominant technology for RF amplifiers and sources in many applications, Vacuum Electron Devices (VED) are viable and practical solutions for megawatt class amplifier systems. Among these devices, the klystron amplifier is at the moment the only commercially available solution for this particular application, where 1 MW is required at a single frequency near 700 MHz. A klystron has the combination of high gain and high efficiency, while being durable and robust. Two klystrons have been built by CPI rated for 1 MW CW at ~700 MHz: one for the Accelerator Production of Tritium at Los Alamos National Lab in the late 1990's, and one for the Brookhaven National Lab in 2006 (Fig. 24). The demonstrated efficiency for these devices was around 65%. For this reason, the klystron is viewed as a relatively mature technology that may be adapted for the RF amplifier necessary to drive the SRF system.

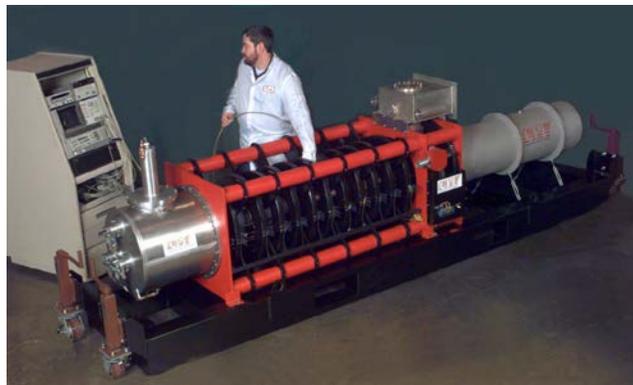

FIG. 24. MW-class, 700 MHz CW commercial klystron manufactured by CPI (model VKP-7952) [41].

Among developing high-power RF amplifier technologies, the multi-beam inductive output tube (MBIOT) may present a realistic alternative. Although IOTs are typically found in signal broadcast systems, they have recently gained ground in the area of accelerator drivers. The

MBIOT has been developed by L3 Electron Devices to provide RF power for the European Spallation Source. To date, it has demonstrated over 1 MW peak power at 704 MHz with 65% RF efficiency [42].

Other VEDs such as magnetron oscillators may be considered for this application. Magnetron technology is as mature as klystrons, and low-weight inexpensive magnetron devices are commercially available which produce 100 kW continuous RF near the required frequency at efficiencies close to 90%. However, a magnetron is an oscillator not an amplifier, and its output frequency and phase are particularly sensitive to the cathode voltage, heater voltage, and reflected power. For this application, a magnetron requires signal feedback and a fast-response control system to produce a stable phase coherent signal necessary to drive an accelerator cavity. In addition, to achieve the required 1 MW, multiple magnetrons must be configured such that they lock together in phase. Phase locking has been demonstrated for magnetrons [43], but for no more than a few devices at moderate (kW) power levels. Although there are many benefits to a magnetron-based RF system, significant development is required.

Although vacuum technology has dominated in the realm of high power RF, solid state has been making rapid advances in this area such that it must be considered for the SRF system. An array of power amplifier modules has been constructed for accelerator applications which can reach power levels near 100 kW [44]. Thales offers a power module array (PMA) system which it claims can produce 250 kW at 600 MHz with 70% DC-to-RF efficiency. There are many advantages to a solid-state based system: it operates at low voltage, requires less cooling than a klystron, and is highly modular. Some amount of development effort is required to produce an array of solid-state power amplifiers with a combiner network capable of generating 1 MW, but the technology must be considered for future iterations of the SRF accelerator system. The modularity of solid state is particularly attractive. For example, if additional cavities were added to the system for higher beam energies, solid state devices may be easily distributed in a way to provide RF for each cavity.

The key performance and operating parameters for the four RF amplifier options are provided in Table V. Some of the parameters are extrapolated to a 1 MW system based on existing devices, which require development. For example, there is no 1 MW solid state array so the number of amplifier units for the existing 65 kW system are multiplied accordingly (with weight and cooling scaled using the advertised 250 kW system from Thales), while maintaining voltage and gain. In reality the DC to RF efficiency will be less since there will be additional combining stages.

TABLE V. Summary of key performance and operating parameters for RF systems

|  | VACUUM ELECTRON DEVICES | | | SOLID STATE |
|---|---|---|---|---|
|  | Klystron | MBIOT | Magnetron | PMA |
| Power per unit (kW) | 1000 | 600 | 100 | 0.7 |
| Number of Units | 1 | 2 | 10 | 1400 |
| DC to RF Efficiency (%) | 50 | 55 | 88* | 70 |
| Voltage (kV) | 95 | 42 | 21 | 0.05 |
| Gain (dB) | 40 | 21 | N/A | 21 |
| Cooling requirement (gpm) | 350 | 250 | 55 | 100 |
| Highest demonstrated CW power (kW) | 1000 | 80** | 100*** | 65 |
| Weight (kg) | 2700 | 1400 | 100 | 1500 |

*Efficiency is for a single device; does not take into account combining losses

**Results are for a single IOT; MBIOTs have demonstrated over 1 MW in pulsed operation

***Results are for a single magnetron; no system of 100 kW magnetrons has been phase-locked to date

**B. Low-level RF and beam diagnostic**

The low-level RF system for the type of accelerator being considered in this study is not particularly challenging, compared to that for particle accelerators for scientific research: a field control amplitude and phase of 1% and 1° should be sufficient as there are no stringent requirements on the beam emittance. A system based on a digital self-excited loop, such as the one designed and built for the CEBAF Upgrade [45], is a possible choice. Two such systems, one for the injector and one for the SRF cavity would be needed. A common RF source or two separate ones, phase locked to one another, would be needed as well. Given the high beam current and therefore low $Q_{ext}$ of the FPCs, the bandwidth is quite large, ~7.8 kHz, making the cavity less sensitive to microphonics.

Beam diagnostic would consist of four sets of beam position monitors, placed before and after each of the focusing solenoids and two beam current monitors, one between the injector and the first solenoid and one after the second solenoid. The beam position monitor could be a water-cooled four-segment aperture. The beam current monitor could be a parametric current transformer. Four small corrector solenoids would be installed before and after each of the two focusing magnets. Interlock signals could be provided by the ion pumps' current, the temperature of the cavity and the reflected power. Increasing the amplitude of the negative DC bias voltage on the injector's grid could be used as a fast way to suppress the electron current, when necessary.

**C. Radiation Shielding**

This section presents an estimate of radiation shielding for the accelerator. At the exit of the linac the electron beam could be used to irradiate the wastewater or flue-gases by either horizontal and vertical scanning. Because the design of the irradiation area at forward angles with respect to the electron beam is not known, this first investigation explores lateral shielding needed to contain a point loss, such as may occur when the electron beam hits a beam pipe at a shallow angle. However, as shown below, the source terms and attenuation length for the forward directions will be similar at this energy (for identical beam loss).

*1. Source term and reference shielding data*

Both source terms and shielding data for different materials are available in Reports 51 [46] and 144 [47] of the National Council on Radiation Protection and Measurements (NCRP). At 1 MeV and 1 MW, the source term for "a thick target of high-Z material" (W, Pb and such) is $2.4\times10^4$ Sv/h at 1 m. It is interesting to note that just around this energy the yield curves for the lateral and forward source term intersect, therefore the dose rate distribution with respect to the incident beam direction will be roughly isotropic. Source term adjustment factors for "low Z" targets are suggested, specifically 0.7 for steel and 0.5 for aluminum.

The most likely source of radiation in the lateral direction is perhaps a full or partial beam loss hitting a beam pipe or a flange under a glancing angle. Experience from CEBAF indicates that beam losses of the order of 1 kW lead to vacuum leaks or other failures within minutes, and more quickly at greater loss rates. It is expected that machine protection interlocks will be set to prevent beam loss orders of magnitudes lower, but a conservative approach suggests designing shielding for a beam loss that could be sustained for long periods or indefinitely, in the absence of active safety devices, such as interlocks, that could possibly fail. We assume here that 0.1 kW is the maximum beam loss that could be sustained for a long time or indefinitely, which represents a fraction of $1\times10^{-4}$ of the total available beam power. The beam pipe or other beamline components will likely be made of steel or aluminum, so the respective source terms will be 1.68 and 1.2 Sv/h at 1 m.

2. *Lateral shielding for chronic beam loss*

Let us assume that a concrete shielding starts at a distance of 1 m from the loss point (beamline) and that dose rates outside the shielding should allow occupancy by non-radiation workers. This is equivalent to limiting the dose accrued over a period of 2000 h of operation (a year of full time occupancy) to 1 mSv, or dose rates lower than $5\times10^{-4}$ mSv/h outside the shielding. The required shielding therefore needs to attenuate by over six order of magnitude, or 6.53 and 6.38 tenth-value layers (TVLs) for the steel and aluminum target, respectively. NCRP provides TVL values for radiation exiting targets forward, at 0°. For 90° it suggests using TVLs at 0° for lower energies given in a graphic appendix. Using this adjustment, the "lateral" TVL value for concrete at our energy is approximately 13 cm. The corresponding thickness of ordinary concrete needed to shield a 0.1 kW beam loss would be ~85 cm and 83 cm for the steel and aluminum pipe, respectively.

3. *Monte Carlo calculations*

Source term and attenuation data present in NCRP reports are based on large sets of experimental data gathered and verified over decades and as such are credible. However, one may wonder how good are the adjustment factors for source terms in low Z materials, and whether the latter may emit softer spectra of photons, in particular at low energies where collision-excitation source of photons may not be negligible compared to bremsstrahlung. We therefore performed a simulation test using the Monte Carlo (MC) code FLUKA [48]. These simulations were performed in a symmetrical (cylindrical) geometry for sake of efficiency. This could slightly enhance the initial source term due to enhancement of the photon flux due to in-scattering within the cylindrical walls. However, considering that scattered photons have lower energies and only photons with the highest energies will penetrate (or generate secondary particles that contribute to the dose outside) thick shielding, this enhancement will have little effect in our case. Assuming that electrons generate a quasi-point source on impact and adjusting for the different distance in the MC simulations, the source terms derived from FLUKA simulations are only slightly lower than the NCRP values, by ~11% for the steel and ~9% for the aluminum target, respectively, as shown in Fig. 25. Attenuation for radiation from the steel target in concrete derived from data in the depth range of 100 to 150 cm yields an estimated TVL of 12.7 cm, which is in good agreement with the 13 cm cited above from NCRP. Attenuation of radiation from the Al target indicates TVL of ~12.3 cm at depths around 100 cm. Approaching depths of 150 cm the attenuation curve deflects slightly upwards, suggesting a TVL ~15.8 cm. This could result from the fact that lower energy photons have been filtered out and only the hardest component remains. However, due to the larger uncertainty of data at this depth in this heavily biased simulation, longer and more careful simulations would be needed to confirm this hypothesis.

While the source term forward would be four orders of magnitude higher for a point source, spatial distribution of the beam over a wide area may lower the source term by one or two orders of magnitude. In practice we could expect having to add a few TVLs in the forward direction. One could guess that perhaps the forward shielding would be ~25% thicker than the lateral.

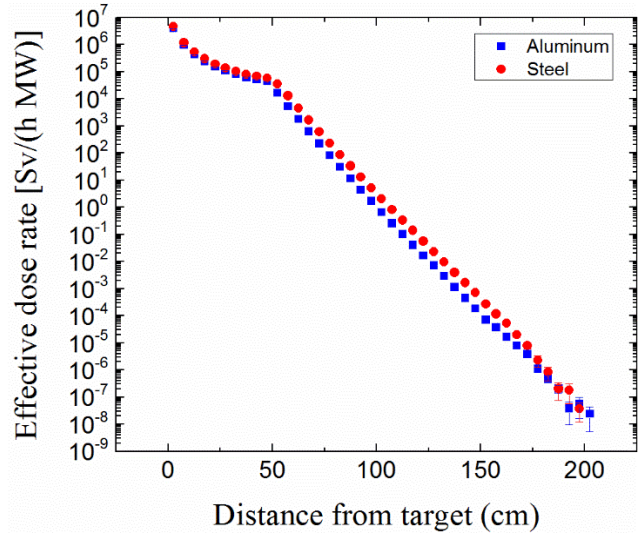

FIG. 25. Effective dose rate as a function of distance from the target computed with FLUKA for stainless steel and aluminum.

## D. Conceptual Layout of Commercial Facility

The commercial environmental accelerator is intended as a standalone module containing all system components that can be directly integrated into existing industrial plants. The layout of such a module with a klystron RF amplifier is illustrated in Fig. 26. The module is compact with dimensions on the order of 10 m in length, 6.5 m in width and 4 m in height, while offering sufficient access for maintenance. Application-specific material handling systems are not shown in Fig. 26. The modules require an external power source, with integration to the industrial plant control and cooling systems for turn-key operations.

In Fig. 26, the shield module containing the cavity and RF amplifier is displayed in dark brown. An electrical enclosure (light gray) with appropriate EMI shielding contains supporting electronic components such as power supplies, capacitor banks, Helium compressors, and diagnostic and control cabinets. The electron beam collector of the klystron requires a large amount of cooling provided by a large industrial chiller separated from the accelerator module.

Multiple additional items require active cooling, requiring a series of smaller chillers, which are placed on the accelerator pad outside of the enclosures to prevent any water leak from damaging sensitive electronics. Air-cooled chillers that do not require external water support are integrated to the facility layout. If a facility water source is provided, the cooling plant footprint can be slightly reduced using smaller water-cooled chillers.

The interior layouts of the facilities with klystron and dual-MBIOTs are presented in Figs. 26a and 26b, respectively. For the facility with the klystron, the beamline is mounted on a platform on top of the klystron to simplify the waveguide routing and minimize losses. The klystron also must be located inside the shield room because of the x-ray radiation generated from its energetic electron beam. For maintenance, access exists along the entire length of the accelerator. Furthermore, sufficient room is left for the large number of utility lines which must reach the equipment from the corner penetration. For the configuration with MBIOTs, the L-shaped electrical enclosure covers the back half of the plant. Since the MBIOTs contain their own shielding, they can be located outside the shield module. A penetration on the chiller side allows access into the shield module for both cooling water, electrical power, and control/diagnostic cables and air handlers, which are located on the roof above the MBIOTs.

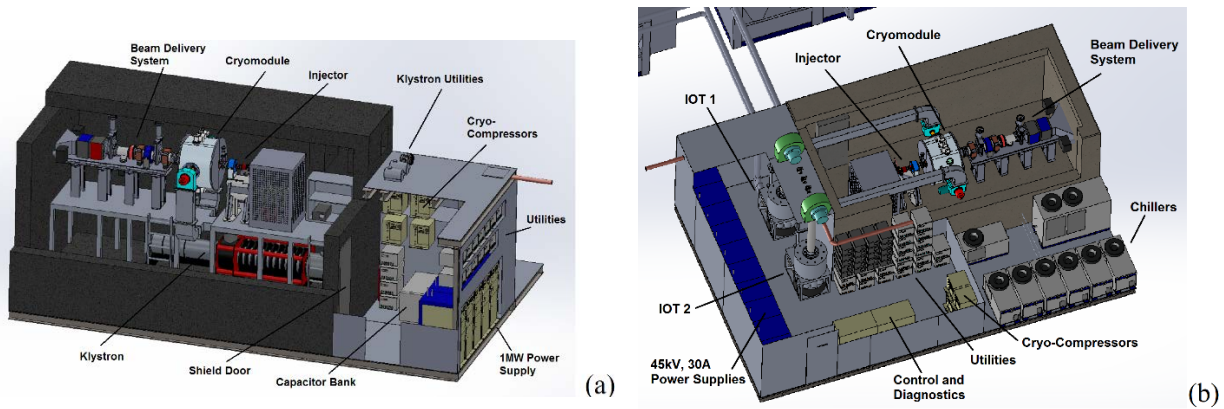

FIG. 26. Interior layouts of the commercial facility with CPI klystron (a) and MBIOTs (b).

### E. Wall-to-Beam Efficiency

The overall wall-to-beam efficiency of the environmental accelerator was estimated at 42%. A simplified energy flow diagram is presented in Fig. 27. Approximately, 2.7 MW of electric power is needed to produce a 950 kW electron beam. The largest demand in electric power comes from the high-power RF amplifiers and, therefore, the overall efficiency strongly depends on the RF amplification efficiency. The RF amplifiers deliver 1 MW of RF power into the cavity with an efficiency of 50% (Table V). Therefore, nearly 2 MW of electric power is required to operate the klystron. In addition, more than 600 kW of water cooling is required to operate the RF amplifiers, the cryocoolers and the beamline components, which necessitates approximately 200 kW of electric power for a facility using exclusively air-cooled water chillers (energy efficiency ratio of ~10).

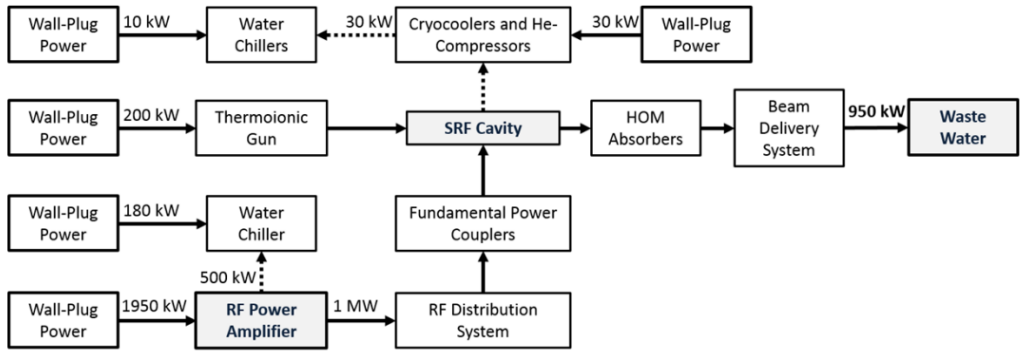

FIG. 27. Simplified energy flow diagram.

### F. Cost Analysis

The calculations of the accelerator capital cost totaling 4.50 M$ are detailed in Table VI. The cost of the RF amplifiers of 3.2 M$ accounts for approximately 70% of the value of the accelerator. It includes the RF power amplifier and auxiliary power supply and control systems, the RF distribution system and dedicated water chillers. The CPI klystron also requires additional

X-ray radiation shielding. The capital cost in Table VI were either provided by or estimated for an industrial production process accordingly to similar projects and experience at Jefferson Lab and AES (e.g., Cu/Nb/Nb$_3$Sn cavity, cryomodule, HOM absorbers, etc.).

TABLE VI. Capital cost of SRF accelerator components

| | |
|---|---:|
| **Total** | **4,500 k$** |
| **RF power amplifier** | **3,200 k$** |
| RF amplifier | 1,100 k$ |
| DC power supply & Controls | 1,900 k$ |
| RF drivers | ~1 k$ |
| RF distribution System | 50 k$ |
| Water chillers | 100 k$ |
| Additional Shielding | 50 k$ |
| **Cryomodule** | **920 k$** |
| Cu/Nb/Nb$_3$Sn Cavity | 160 k$ |
| FPCs | 282 k$ |
| HOM Absorbers | 63.8 k$ |
| Vacuum Vessel | 40 k$ |
| Thermal Shields | 10 k$ |
| Magnetic Shields | 40 k$ |
| Gate Valves | 145.4 k$ |
| Cryocoolers | 168.8 k$ |
| Vacuum Pump | 10 k$ |
| **Injector** | **217 k$** |
| DC power supply + Control | 65 k$ |
| 200W RF amplifier | 9 k$ |
| RF Control & Interlock | 55 k$ |
| Thermionic gun | 23.5 k$ |
| Magnets | 36.5 k$ |
| Vacuum pump & Flanges | 28 k$ |
| **Beam Delivery System** | **125 k$** |
| Magnets | 23.8 k$ |
| Raster Magnet & Control | 38 k$ |
| Scanning Horn & Ti window | 36.5 k$ |
| Horn Water chillers | 16.7 k$ |
| Vacuum Pump | 10 k$ |
| **Beam Diagnostics & Controls** | **38 k$** |
| Beam position monitors | 8 k$ |
| Beam viewers | 10 k$ |
| Controllers & Acquisition | 20 k$ |

The cost of infrastructure and installation is estimated at 2.75 M$ (Table VII). The cryomodule assembly requires a special facility including a clean room and specific RF conditioning, tooling and testing equipment as well as the participation of a team of four technicians and three engineers during a 10 man-week effort. The installation cost of the cryomodule of 1.0 M$ (Table VII) is expected to reduce by 20% at production scale. 25% of the cost of infrastructure and installation is added to account for transportation, taxes and insurance fees.

TABLE VII. Cost of infrastructure and installation

| | |
|---|---:|
| Total | 2,750 k$ |
| Cryomodule assembly and installation (labor) | 1,000 k$ |
| Low-level RF system assembly and installation (labor) | 100 k$ |
| Civil work | 300 k$ |
| Raw materials (walls, shield, etc) | 500 k$ |
| Radiation monitoring & Safety system | 50 k$ |
| Utilities | 100 k$ |
| Handling system installation | 200 k$ |
| Others (transport, tax, insurance) | 500 k$ |

The estimated accelerator capital expenditure (CAPEX) which includes the costs of the accelerator, infrastructure and installation is approximately 7.25 M$, which corresponds to a value of $7.6 per watt of beam power. In Table VIII, the operating cost was computed according to the following assumptions:

i. The accelerator is operated for 8,000 hours per year, corresponding to an average daily usage of approximately 22 h and allowing for more than 700 hours per year of routine maintenance;

ii. Annual cost of maintenance is evaluated at 2% of the capital value of the system;

iii. The accelerator operates as a turnkey system, fully integrated to industrial facility systems and controls;

iv. No full-time, dedicated operator is required to operate and monitor the system. The accelerator will be one of the various systems to be controlled by plant operators already working at the facility.

v. Cost of electric power: $0.07/kWh;

vi. No external water supply required (facility equipped with air-cooled water chillers only);

vii. Dose deposition efficiency: $\pi_e = 60\%$.

Amortization was calculated for a 15-year, 8%-rate period and a 20% initial investment of 1.45 M$. For both RF configurations (klystron and MBIOT), amortization reaches 670 k$ per year, which reduces to approximately $85 per hour of operation. The electric power consumption

is in the order of 2.75 MW for both systems, which cost $159.2/hr. The cost of maintenance of 140 k$ per year reduces to $17.5 per hour of operation. The total cost operation is of $262/hr with a klystron and $251.2/hr with MBIOTs. In summary, it is anticipated that 61% of the cost of operation will be associated to electric consumption, 32% for amortization and the remaining 7% for maintenance (Fig. 28).

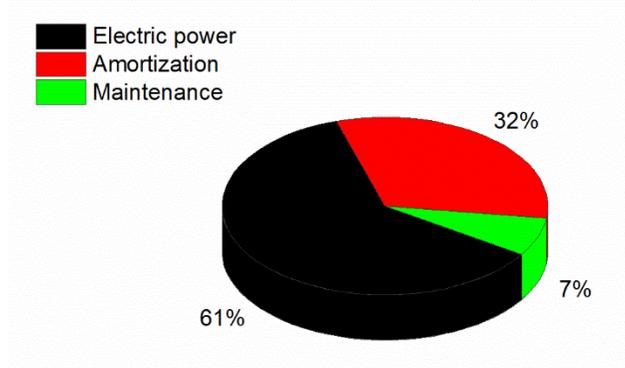

FIG. 28. Breakdown of the processing cost.

TABLE VIII. Estimated costs of operation and material processing.

| Capital Cost | |
|---|---|
| SRF Accelerator | $4,500,000 |
| Infrastructure | $2,750,000 |
| Investment (20%) | $1,450,000 |
| Amortization (15 yrs @8%) | $677,611/yr |
| Operation Expenses | |
| Power ($/hr) | $159.2/hr |
| Maintenance | $145,000/yr |
| Total Op. Cost | $177.3/hr |
| | $1,418,367/yr |
| Total Cost of Operation | $262/hr |
| | ($2,095,978/yr) |
| Processing Cost (¢/ton/kGy) | 12.75 |

The cost of cooling water is not included in the calculations of Table VIII. Air-cooled water chillers do not require an external water supply. Water-cooled chillers are more efficient than air-cooled systems, but more expensive and require external water supplies. Overall, it is estimated that the cost of operation remains the same with either air-cooled or water-cooled chillers.

The processing cost, denoted $C_{process}$, is defined as the cost for deposition of a dose of 1 kGy into a ton of materials. Equivalently, since 1 Gy = 1 J/kg, $C_{process}$ is the cost for deposition of 1 MJ of radiation. The processing cost can be expressed as:

$$C_{process}(¢/ton/kGy) = 277.8\ C_{oper}(\$/hr)/\pi_e P_{beam}(kW), \qquad (7)$$

where $C_{oper}$ is the total hourly cost of operation, $P_{beam}$ the electron beam power, and $\pi_e$ the dose deposition efficiency. The estimated processing cost is approximately 12.75 ¢/ton/kGy with a klystron (Table VIII).

For a specific application such as wastewater treatment, the processing cost and the daily processed volumes ultimately depends on the required dose deposition (Table IX). For a 1 MW SRF accelerator delivering a 4 kGy dose into wastewater, which is necessary for killing >99% bacteria [49, 50], the processing cost would be approximately $0.51/m³ ($1.930/kgal) and more than 11,000 m³ (~3 Mgal/day) of wastewater could be processed daily. A lower dose requirement of 1 kGy results in a fourfold daily processed water volume capability of 45,000 m³ (11.9 Mgal/day), requiring the handling of a large wastewater flow rate of ~9,000 gpm. It demonstrates that the operating cost of a 1 MW SRF accelerator facility could be very cost-effective for industrial wastewater treatment at small scale (<10 Mgal/day).

For application in the treatment of flue gases, operation of the pilot plant in Poland showed that a 1 MW accelerator should be capable of treating the flue gases from a 100 MW coal power plant. Considering an average capacity factor of 60%, corresponding to ~5300 h per year, the processing cost would be ~16 $/kW, which is quite competitive with wet-scrubbing technologies. The processing cost could in fact be even lower given that a large fraction of the cost is electricity and that the power plant itself would provide it.

TABLE IX. Wastewater treatment cost and daily processing capability for various dose requirements

|  | REQUIRED DOSE DEPOSITION | | |
| --- | --- | --- | --- |
|  | 1 kGy | 4 kGy | 10 kGy |
| Cost of processing 1 m$^3$ | $0.1275 | $0.510 | $1.275 |
| Cost of processing 1 Mgal | $482 | $1,930 | $4,825 |
| Daily Processed Volume | 45,000 m$^3$ (11.9 Mgal) | 11,250 m$^3$ (3.0 Mgal) | 4,500 m$^3$ (1.19 Mgal) |
| Required Flow Rate (gpm) | 9,050 | 2,260 | 905 |
| Comments [41, 42] | Color, Odor, Coliform bacteria removal | Kill >99% of bacteria | Inactivate some radiation resistant organisms |

## G. Uncertainty of Processing Cost

The validity of the economic assessment relies on proper estimations of critical design parameters. The sensitivity of the cost analysis to key quantities is analyzed in this section.

Figure 29 shows the dependence of the processing cost to the RF efficiency and the annual operation time.

The cost of electric power accounts for 60% of the processing cost, which makes the processing cost strongly dependent on the overall efficiency of the system. If RF power is not efficiently generated and harnessed through the cavity, the processing cost increases rapidly. Figure 29a shows that the processing cost increases by 20% for an RF power efficiency of 50%. Reciprocally, in the upper limit of 90% of RF power efficiency, as theoretically conceivable with magnetrons, the processing cost is predicted to decrease by more than 20% to less than 10 ¢/ton/kGy. Taking also into account the low capital cost of magnetrons, this could motivate the development of a high-energy CW magnetron-based RF system.

Estimating the cost of maintenance is challenging. Routine maintenance is expected to include frequent replacement of the titanium window and periodic servicing of the RF systems, chillers and cryocoolers. As such, the estimated annual maintenance cost of 2% of the accelerator CAPEX may be a conservative estimate in absence of unexpected system failure. If the system fails and requires a long downtime, the impact of reduced operation time can be severe on the processing cost (Fig. 29b). In a scenario where the annual running time is reduced by 25% to 6,000 hrs, the processing cost is anticipated to increase by only 13% to 14.4 ¢/ton/kGy. Longer downtime, where the system can only operate for 4,000 hours, results in a 40% higher processing cost of 17.75 ¢/ton/kGy. Such an increase points towards the development of a solid-state system, where there is no single point of failure, and maintenance time is expected to be less than a vacuum device.

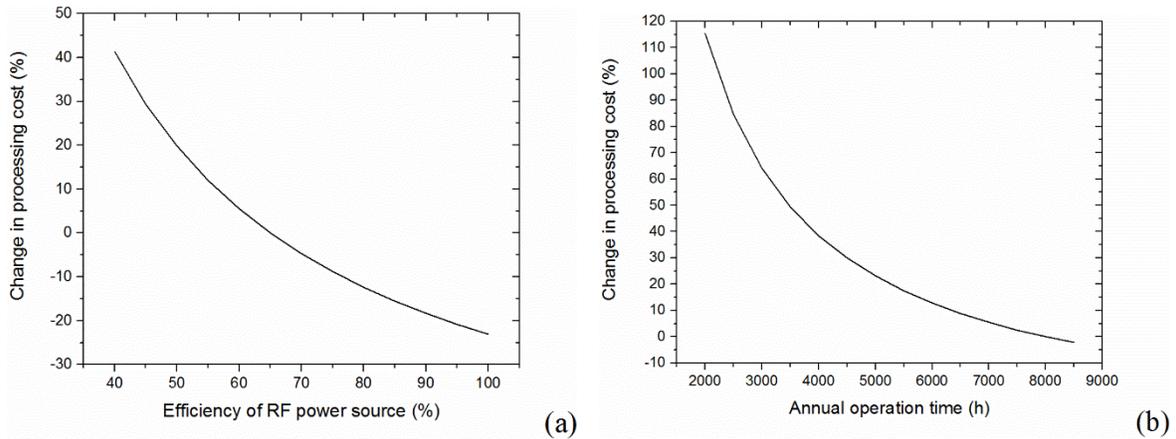

FIG. 29. Processing cost dependence on RF driver efficiency (a) and annual operation time (b).

## V. DISCUSSION

RF power is the dominant cost driver for high-power accelerators envisioned for environmental remediation. Whereas present commercial vacuum electron devices (klystrons or MBIOTs) are a viable option for this application, further R&D in alternative sources such as magnetrons or solid-state amplifiers is desirable to keep improving the overall efficiency and reducing the cost of the accelerator.

Although this study shows that a cryogen-free, 1 MeV, 1 MW SRF accelerator is technically feasible and economically viable, it might be worthwhile to compare it with a normal-conducting configuration. For example, a 1497 MHz slot-coupled normal conducting capture cavity was recently designed for efficient CW operation to accelerate a 0.4 mA electron beam from 130 keV to 510 keV [51]. Scaling such cavity to 750 MHz and to about ten cells to accelerate the beam to ~1 MeV would result in ~13 kW of power loss in the copper walls, resulting in a similar wall-plug power as that required to operate the four cryocoolers in the SRF design. However, the cavity length would be ~2 m, that is, nearly twice that of the SRF cryomodule, and the iris radius being ~1 cm could lead to significant beam scraping, which could damage the cavity given the high beam power. A different normal conducting accelerating structure recently demonstrated the possibility to accelerate an electron beam from 15 kV to ~1.2 MeV, with a beam current up to 25 kW, in a compact CW RF accelerator for industrial applications [52, 53]. The structure resonates at 2.45 GHz, it is ~1.3 m long and it dissipates ~20 kW in the copper walls. Issues related to beam scraping apply to this design as well when considering a beam power of 1 MW.

The attractiveness of the SRF option is the possibility of maintaining a high efficiency and low footprint when scaling-up the design to achieve a beam energy of 10 MeV, still with a 1 MW power, which would expand the accelerator application to the treatment of sludge and medical waste. The higher beam energy could be achieved, for example, with a low-$\beta$ two-cell cavity after the injector, with an energy gain of ~2 MeV, and two $\beta=1$ three-cell cavities each providing an energy gain of ~4 MeV. Given the reduced beam current, a single FPC could be used for each cavity and adequate damping should be achievable in three-cell structures. The three cavities could be packaged in a single cryomodule about 3 m long. In case where multiple cavities are needed, adopting the standard cavity cooling with liquid Helium at 4 K becomes cost-competitive instead of using an increasing number of cryocoolers. Compact liquid Helium refrigerators/liquefiers with cooling power of up 100 W at 4.6 K are commercially available [54], the systems with turbine-based expanders cost ~1 M$ and have long maintenance intervals (~30,000 h), whereas systems with piston–based expanders are about half of the cost but they require a 2-3 day maintenance shutdown every ~5,000 h [55].

Some technical challenges related to the accelerator design presented in this study will have to be demonstrated with prototypes: the detailed design of the shape of the electrodes of the thermionic gun would need to be finalized and a prototype should be built. Similarly, conduction cooling of an elliptical-shaped SRF cavity at peak surface fields similar to those envisioned for this accelerator should be demonstrated on a prototype cavity. The surface resistance of cavities made of dissimilar metals is strongly affected by the temperature uniformity at the transition temperature. Since the cooling rate of cryocoolers is significantly lower than what is typically achieved by direct cooling with cryogenic fluids, low thermal gradients along the cavity can be expected by using cryocoolers but will have to be verified in a prototype cryomodule. Deposition of copper or other high-thermal conductivity material onto a Nb cavity with good adhesion and good thermal properties should also be demonstrated. Currently, high-temperature vacuum furnaces used for the development of $Nb_3Sn$ and that have sufficiently large hot zone to allow coating of a 750 MHz single-cell cavity are available both at Jefferson Lab and Fermilab. High-temperature vacuum furnaces are also available at companies specialized in cavity manufacturing and the adaptation of such furnaces for the $Nb_3Sn$ coating process is not expected to be a major investment.

Further engineering design followed by validation with prototypes should be devoted to the beam rastering system and beam exit window, in order to optimize the beam distribution over the window area and to minimize the power density at the window. Lifetime tests of the beam exit window will be important aspects of the accelerator development.

Beam losses need to be absolutely avoided as the high-power beam could cause significant damage to the accelerator and produce large amounts of radiation. Even a very small fraction of beam loss inside the cryomodule can be very deleterious as it could cause a thermal quench of the SRF cavity. The beam transport simulations showed that no particle is lost on the beam wall chamber and the choice of large diameter beamline tubes should certainly help achieving this in practice.

## VI. CONCLUSIONS

The accelerator design discussed in this article shows that it is technically feasible to envision a 1 MW, 1 MeV CW SRF electron accelerator for the treatment of flue gases and wastewater. Advances in SRF and cryogenic technologies make it feasible to use a cryogen-free cryomodule with a single SRF cavity to provide most of the energy. The cost analysis show that the accelerator design discussed in this article would be a cost-competitive solution to the problem of wastewater and flue gas treatment.

Given the encouraging outcomes of this design study, additional R&D funding would enable to demonstrate the performance of prototype components and to address some of the technical challenges, eventually leading to a full-scale demonstration of the accelerator.

## ACKNOWLDEGMENTS

We would like to acknowledge Tom Powers and Curt Hovater, Jefferson Lab, for helpful discussions on the low-level RF, Kevin Jordan and Matt Poelker, Jefferson Lab, for helpful discussions about beam diagnostics, George Karashvili, Jefferson Lab, for helping with the FLUKA simulation. This manuscript has been authored by Jefferson Science Associates, LLC under U.S. DOE Contract No. DE-AC05-06OR23177. Funding for this work has been provided by the DOE High-Energy Physics Accelerator Stewardship program. The U.S. Government retains a non-exclusive, paid-up, irrevocable, world-wide license to publish or reproduce this manuscript for U.S. Government purposes.